\documentclass[twocolumn,tightenlines,aps,preprintnumbers,superscriptaddress,showpacs]{revtex4}
\usepackage{graphicx}
\usepackage{amssymb}

\newcommand{\lgtr}{\scriptscriptstyle\lessgtr}
\newcommand{\lapthree}{\Delta^{\scriptscriptstyle{(3)}}}

\begin{document}
\preprint{LA-UR-04-3183}
\preprint{DO-TH-04/06}
\title{Characterizing Inflationary Perturbations: The Uniform Approximation}

\author{Salman Habib}
\affiliation{T-8, Theoretical Division, MS
B285, Los Alamos National Laboratory, Los Alamos, 
New Mexico 87545, USA}

\author{Andreas Heinen}
\affiliation{Institut f\"ur Physik, Universit\"at Dortmund,
D - 44221 Dortmund, Germany}

\author{Katrin Heitmann}
\affiliation{ISR-1, ISR-Division, MS
D436, Los Alamos National Laboratory, Los Alamos, 
New Mexico 87545, USA}

\author{Gerard Jungman}
\affiliation{T-6, Theoretical Division, MS
B227, Los Alamos National Laboratory, Los Alamos, 
New Mexico 87545, USA}

\author{Carmen Molina-Par\'{\i}s} \affiliation{Department of Applied 
Mathematics, University of Leeds, Leeds LS2 9JT, UK}

\date{\today}

\begin{abstract}
    
The spectrum of primordial fluctuations from inflation can be obtained
using a mathematically controlled, and systematically extendable,
uniform approximation.  Closed-form expressions for power spectra and
spectral indices may be found without making explicit slow-roll
assumptions. Here we provide details of our previous calculations,
extend the results beyond leading order in the approximation, and
derive general error bounds for power spectra and spectral indices.
Already at next-to-leading order, the errors in calculating the power
spectrum are less than a per cent.  This meets the accuracy
requirement for interpreting next-generation CMB observations.

\end{abstract}


\maketitle

\section{Introduction}

Recent observations of the cosmic microwave background
(CMB)~\cite{cmbobs,rev} and of the galaxy distribution in real and
redshift space~\cite{lssobs} have dramatically improved our knowledge
of the large-scale Universe. Results from these observations have been
largely consistent with the inflationary paradigm of cosmology, since
it appears that the Universe is at critical density and that
the observed structure in the Universe arose from
the gravitational collapse of adiabatic, Gaussian, and
nearly-scale-invariant primordial density fluctuations, a key
prediction of simple inflation models.

The idea that an inflationary expansion, aside from solving
cosmological puzzles such as the horizon and flatness problems, also
provides a natural mechanism for the generation of primordial
fluctuations was realized around the time the model was first
proposed~\cite{infpertorig}. Early estimates already established the
essential adiabatic, Gaussian, and nearly-scale invariant nature of
the perturbation spectrum. However, these estimates were rather crude
and their true value lay in providing very significant qualitative
guidance rather than accurate quantitative information. As the
prospect of testing inflation and even distinguishing between
individual models from precision CMB observations became more
realistic, attention was increasingly focused on improving the
understanding and predictive control of the theory underlying the
generation of fluctuations. The development of the gauge invariant
treatment of cosmological
perturbations~\cite{givb,givs,ivrevs1,ivrevs2} and its application to
inflation~\cite{ivrevs2,givinf,givvfm,givdhl} was of great help in
clarifying and systematizing the calculations involved.

The simple nature of the perturbation spectrum characteristic of
inflation arose from models where inflation was caused by the dynamics
of a single scalar field in a relatively featureless potential,
evolving in an effectively friction-dominated `slow-roll'
regime~\cite{LLKCBA,slowrev}. It was realized that in more complex
inflation models (multiple fields, nontrivial potentials) it is
possible to engineer the spectra in a variety of ways including
violation of scale-invariance and introduction of isocurvature
fluctuations. However, data from present-day observations do not
demand the consideration of dynamically exotic inflation models.

Inflation predicts a spectrum of metric perturbations in the scalar
(density) and tensor (gravitational wave) sectors, the vector
component being naturally suppressed. Both scalar and tensor
perturbations cause anisotropy in the CMB
temperature~\cite{anisotropy}. Scalar fluctuations seed formation of
large-scale structure in the Universe, while the tensor perturbations
lead to a stochastic gravitational wave background~\cite{gravbg}. In
addition, tensor modes cause a potentially observable polarization of
the CMB, a target for next-generation CMB
observations~\cite{polar}. Given the variety, precision, and volume of
data now and soon to be available from the various probes of
primordial fluctuations, it is important to provide controlled
theoretical analyses for various classes of inflation models.

One obvious application of such an analysis lies in comparing
theoretical predictions to observations in order to test specific
inflation models.  In addition, there is the prospect of obtaining
information about the slow-roll potential via a controlled inverse
analysis of observational data, the so-called `reconstruction'
program~\cite{recon} (for a review see~\cite{LLKCBA}). Finally, there
is also the more general question of whether the inflationary paradigm
itself can be tested from observations. Regarding this question,
recent activity has focused on identifying classes of inflation models
distinguished by their differing observational signatures~\cite{HT}
rather than trying to reconstruct a specific inflationary potential.
A recent discussion of parameter estimation using input from inflation
calculations can be found in Ref.~\cite{paraminf}.

To expand further on the above, we note that while inflation is the
most compelling cosmological paradigm, it is not without its
competitors~\cite{other}.  None of the alternative scenarios,
including structure formation via topological defects (definitively
ruled out by observations), or string-inspired models such as the
ekpyrotic and cyclic Universe scenarios, as yet may be viewed as
offering convincing competition. Nevertheless, an important question
arises in the comparison of predictions particular to generic
models of inflation to predictions one might expect on
purely general grounds or from other models. A good example of this
is provided by the spectral index $n_S$ characterizing the scalar
perturbation spectrum.  The Harrison-Zeldovich choice
$n_S=1$~\cite{hz} is aesthetically natural, independent
of inflation.  A high-precision observational detection of $n_S$
provides a good target for inflation models if $n_S$ is indeed
slightly different from unity.  (Clearly, an intuitive argument such as
that of Harrison and Zeldovich cannot be refined further without
a corresponding theoretical framework.)  Related to this,
another important point is that inflation predictions for the
scalar and tensor spectra are not mutually independent.  Future
measurements of the CMB can provide tests of this dependency, thus it
is important to state precisely the appropriate
`consistency relations'~\cite{cons} and the accuracy to which they can
be calculated for a range of inflation models.

The equations governing the evolution of the scalar and tensor
perturbations can be solved numerically. For specific models, it can
scarcely be argued that an approximate analytical result is
competitive with a mode-by-mode numerical solution of the
equations. Nevertheless analytic results can be extremely powerful in
providing generic results valid for many classes of models.  Accurate
analytic results also greatly reduce the computational
requirements for the reconstruction program.
Therefore some effort has concentrated
on improving theoretical predictions by
sharpening the slow-roll analysis. In this regard, approximations
based on certain slow-roll assumptions have been criticized on
multiple fronts. It has been argued that error control is not adequate
and that they are not systematically improvable~\cite{wms,ms}. There
have been some efforts to alleviate this
situation~\cite{srimprov,schwarzetal}, however, they are based on
restrictive assumptions (e.g., $n_S\simeq 1$) and often lead to
complicated mathematical formulations.

Recently, we presented an analysis of the inflationary perturbation
spectrum for single-field models based on applying a uniform
approximation to the relevant mode equations~\cite{hhjm}.  This
analysis goes beyond the slow-roll approximation, does not make
restrictive assumptions, has definite error bounds, and is
systematically improvable. In this paper we implement the method
beyond leading order and provide general error bounds on power spectra
and spectral indices. While our general expressions for the spectral
indices are nonlocal, we can derive local expressions for $n_S$ and
$n_T$ by employing a further approximation. We also demonstrate that
by Taylor-expansion of our local results, expressions of the slow-roll
type can easily be obtained. Finally, in order to demonstrate the
accuracy of the uniform approximation explicitly, we discuss its
application to an exactly solvable inflation model.

The organization of the paper is as follows: Section~\ref{background}
provides the essential background to understanding the problem of
computing the power spectrum, Section~\ref{uniapproximation} explains
the uniform approximation, Section~\ref{lead} gives the leading order
results, Section~\ref{nexttolead} provides the results at next order,
and Section~\ref{LOCAPP} contains the results of Section~\ref{lead}
simplified to a local form. Here we also encapsulate the conventional
slow-roll results and derive the analogous results from the uniform
approximation. Section~\ref{special} considers a special example which
possesses an exact solution and uses this to demonstrate the
characteristic features of the uniform approximation. In
Section~\ref{besselcom} we comment briefly on Bessel function
approximants.  Section~\ref{conclusion} concludes with a discussion of
the results obtained and an outline of future work.
Appendix~\ref{backg} provides a short list of the relevant definitions
and equations for linearized perturbations in an FRW universe.
Appendix~\ref{error} contains the technical details behind the error
formulae of the uniform approximation.

\section{Background}
\label{background}

The cosmic microwave background and the large-scale distribution of
matter both provide information on the primordial perturbation
spectrum, albeit `processed' by physics during the radiation and
matter-dominated phases. Extraction of information from large scale
structure observations is complicated by the nonlinear effects of the
gravitational instability and various sources of observational bias.
Fortunately, the physics of the photon distribution is very different
from that of the mass distribution: As a consequence of radiation
pressure, the CMB is very uniform and fluctuations in it can be
treated adequately by a linearized analysis.

The generation of perturbations during inflation is due to the
amplification of quantum vacuum fluctuations by the dynamics of the
background spacetime. Of course, the post-inflationary epoch
must also be understood in order to make a connection with observations.
A serious treatment of the dynamics of the
post-inflationary phase is necessarily quite complicated as various
effects such as reheating, particle decay, etc. have to be taken
into account.
Fortunately, it turns out that an accurate accounting
of much of this physics is not required in order to make inflation
predictions for CMB anisotropies and the large-scale distribution of
matter.

The modern treatment of fluctuations generated by inflation is in turn
based on the gauge-invariant treatment of linearized fluctuations in
the metric and field quantities
(Appendix~\ref{backg})~\cite{givb,givs,ivrevs1,ivrevs2}).  A
particularly convenient quantity for characterizing the perturbations
is the intrinsic curvature perturbation of the comoving
hypersurfaces~\cite{givdhl}, $\zeta\equiv u/z$, where $u\equiv a
(\delta\phi + \phi'A/h)$ and $z\equiv a\phi'/h$, with $a$, the scale
factor, the prime denoting a derivative with respect to conformal
time, $h\equiv a'/a$, $\delta\phi$ the perturbation in the homogeneous
background scalar field, $\phi$, and $A$, a quantity characterizing a
perturbation of the background metric (details are given in
Appendix~\ref{backg}).  As shown in Appendix~\ref{backg},
the gauge-invariant quantity, $u$, satisfies
the dynamical equation
\begin{equation}
u''-\Delta u - {z''\over z}u=0.
\label{uevol}
\end{equation}
It follows immediately that $\zeta$ is approximately constant in the
long wavelength limit $k\rightarrow 0$.  This is true during the
inflationary phase as well as in the post-reheating era.  Moreover,
the Einstein equations can be used to connect the gravitational
potential $\Phi_A$ and $\zeta$ so that a computation of the power
spectrum of $\zeta$ provides all the information needed (aside from
the transfer functions) to extract the temperature anisotropy of the
CMB. Details of this procedure can be found in
Refs.~\cite{ivrevs2,ms,ms2}.  We will return to some aspects of this
analysis below.

The calculation of the relevant power spectra involves a computation
of the two-point functions for the appropriate quantum operators,
e.g., 
\begin{equation}
\langle 0|\hat{u}(\eta,{\bf x})\hat{u}(\eta,{\bf x}+{\bf
r})|0\rangle=\int_0^{\infty} {dk\over k} {\sin kr\over kr} P_u(\eta,k),
\label{pspectra}
\end{equation}
the operator $\hat{u}$ being written as
\begin{equation}
\hat{u}(x) = \int {d^3k\over (2\pi)^{3/2}}\left[\hat{a}_k u_k(\eta)
\hbox{e}^{i{\bf k}\cdot{\bf x}} + \hat{a}_k^{\dagger}
u_k^*(\eta)\hbox{e}^{-i{\bf k}\cdot{\bf x}}\right] ,
\label{uexpand}
\end{equation}
where $\hat{a}_k,~\hat{a}_k^{\dagger}$ are annihilation and creation
operators, respectively, such that $[\hat{a}_k,\hat{a}_{k'}^{\dagger}]
= \delta_{kk'}$, and defining the vacuum state $\hat{a}_k|0\rangle =
0$ $\forall k$. The complex amplitude $u_k(\eta)$ satisfies
\begin{equation}
u_k^{\prime\prime}+\left(k^2 -{z^{\prime\prime}\over z}\right)u_k=0. 
\label{mode}
\end{equation}
Solving Eqn. (\ref{mode}) is the fundamental problem in determining
the primordial power spectrum $P_u$ (or $P_{\zeta}$). The
corresponding mode equation for tensor perturbations is given by
\begin{equation}
v_k^{\prime\prime}+\left(k^2 -{a^{\prime\prime}\over a}\right)v_k=0. 
\label{modev}
\end{equation}

Equations (\ref{mode}) and (\ref{modev}) have the mathematical form of
Schr\"{o}dinger equations. A simple approach to analytical
approximation of Eqns. (\ref{mode}) and (\ref{modev}) relies on the fact
that exact solutions exist in the limits $k^2 \gg
\left|z^{\prime\prime}/z\right|$, $\left|a^{\prime\prime}/a\right|$
(short wavelength) and $k^2\ll \left|z^{\prime\prime}/z\right|$,
$\left|a^{\prime\prime}/a\right|$ (long wavelength) or, as will be
made more explicit below, as $-k\eta\rightarrow \infty$ and
$k\eta\rightarrow 0^-$.  For scalar perturbations,
\begin{eqnarray}
u_k&\rightarrow&{1\over \sqrt{2k}}e^{-ik\eta}~~\left(k^2 \gg
\left|z^{\prime\prime}/z\right|,~-k\eta\rightarrow \infty\right),
\label{asymp1}\\    
u_k&\rightarrow&A_k z~~\left(k^2\ll\left|z^{\prime\prime}/z\right|\right).
\label{asymp2}
\end{eqnarray}
Here, the short wavelength solution corresponds to the choice of an
adiabatic vacuum for modes on length scales much smaller than the
scale set by the curvature. The long wavelength solutions correspond
to the growing mode on scales much larger than the Hubble
length. (Analogous solutions exist for gravitational wave
perturbations.)

\medskip

\begin{figure}[h]
\leavevmode\includegraphics[width=8.2cm]{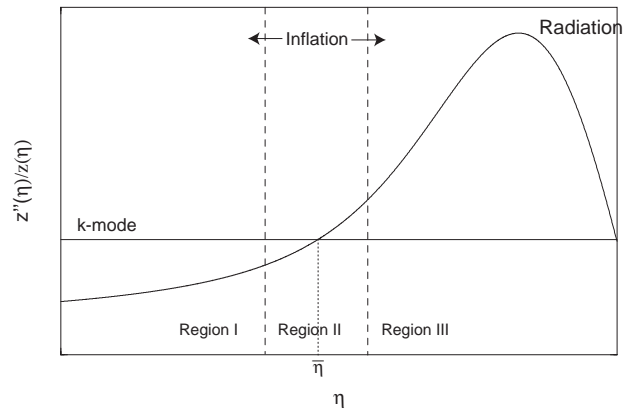}
\caption{Sketch of the potential barrier for density perturbations.
The vertical dashed lines delineate the three different regions where the
solution for $u_k$ is investigated as explained in the text.
$\bar\eta$ marks the point where the $k$-mode enters into the
potential barrier, the turning point. Inflation lasts
over a wide $\eta$-range, extending Region II as indicated by the
arrows.  In the far right of Region III radiation dominates.}
\label{fig1}
\end{figure}

The long wavelength solution for $\zeta_k=u_k/z$ is just the
($k$-dependent) constant $A_k$ of Eqn.~(\ref{asymp2}). In order to
determine the corresponding power spectrum $P_{\zeta}(k)\equiv
P_S(k)$, the main task is to fix the unknown constant $A_k$ by
connecting the two asymptotic solutions. We sketch the situation in
Fig.~\ref{fig1}.  The large-$k$ regime lies in region I and the
small-$k$ regime lies in region III. The two asymptotic solutions may
be connected either by a matching procedure performed in region II,
typical of the slow-roll class of approximations discussed further
below, or, as performed here, by constructing a global interpolating
solution. Once $A_k$ is determined, the power spectrum for $\zeta$ can
be computed in the regime ($k\rightarrow 0$) of actual interest,
\begin{equation}
P_{S}(k)={k^3\over 2\pi^2}\left|A_k\right|^2, 
\end{equation} 
where the index $S$ indicates scalar perturbations. 

\section{The Uniform Approximation}
\label{uniapproximation}

We begin our discussion of the uniform approximation by making the
substitution 
\begin{equation}
{z^{\prime\prime}\over z}\equiv{1\over
\eta^2}{\cal{C}}^2(\eta)
\label{defcc}
\end{equation}
in Eqn. (\ref{mode}), yielding
\begin{equation}
u_k^{\prime\prime}(\eta)+\left[k^2-\frac{{\cal{C}}^2(\eta)}{\eta^2}
\right]u_k(\eta)=0.
\label{modeua}
\end{equation}
A similar substitution can be made for the case of tensor
perturbations.

If, in Eqn. (\ref{modeua}), we make the assumption that ${\cal{C}}$ is
constant, (which is the case for power law inflation, see Section VII)
then an exact solution in terms of Bessel functions is immediate. The
problem is to solve the equation when ${\cal{C}}$ is not constant but
is a slowly varying function of time. Our aim is to do this without
being forced to state ${\cal{C}}(\eta)$ in any real detail. The method
we will use is the technique of uniform approximation as presented by
Olver~\cite{Olver}. We now provide a brief summary of this treatment
for the differential equation of interest (\ref{modeua}).

The differential equation we wish to solve is of the general form
\begin{equation}\label{general}
{d^2 u\over d\eta^2}=\left[b^2 g(\eta)+q(\eta)\right]u.
\end{equation}
Depending on the behavior of $b^2 g(\eta)+q(\eta)$ this equation has
different approximating solutions.  In the case that $ g(\bar\eta)=0$
at the point $\bar\eta$, so $\bar\eta$ is a turning point,
the solution is expressed in terms of Airy functions; if
$g(\eta)$ has a pole of order $n\geq 2$ the Liouville-Green (LG)
approximation must be employed~\cite{Olver}. The Liouville-Green
approximation is essentially equivalent to the
familiar Wentzel-Kramers-Brillouin (WKB) approximation; for an
application of WKB to computing perturbations from inflation, see
Ref.~\cite{ms03}. In our case the relevant function is
$-k^2+{\cal{C}}^2(\eta)/\eta^2$; we have a turning point which will
depend on the explicit form of $g(\eta)$ and a pole of order $2$ in
the limit $\eta\rightarrow 0^{-}$.  Therefore, we have an Airy solution
around the turning point which goes over to an LG solution for
conformal time approaching zero. As will be shown later, this
approximation fits the exact behavior of the solution accurately.  For
the two approximating solutions (Airy and LG), the convergence
criteria are established in the following way.  Assume that we
have a pole of order $n\geq 2$ at a finite endpoint $a_2$ and
that $g(\eta)$ and $q(\eta)$ are meromorphic functions of
the form
\begin{eqnarray}
g(\eta)&=&{1\over (a_2-\eta)^2}\sum_{s=0}^{\infty}
f_s(a_2-\eta)^s,\nonumber\\ 
q(\eta)&=&{1\over (a_2-\eta)^2}\sum_{s=0}^{\infty} g_s(a_2-\eta)^s.
\end{eqnarray}
Then the error control function (see Appendix~\ref{error}) converges if
$g_0=-1/4$. For a detailed proof of this fact, see Ref.~\cite{Olver}. For
the error control criteria to be satisfied in our case, we must
make the choices
\begin{eqnarray}
g(\eta)&=&{1\over \eta^2}\left[{\cal{C}}^2(\eta) +{1\over
4}\right]-k^2,\nonumber\\ 
&\equiv& {\nu_S^2(\eta)\over \eta^2} \label{nu}-k^2,\\
q(\eta)&=&-{1\over 4\eta^2}.
\end{eqnarray}
The final form of Eqn.~(\ref{mode}) then becomes
\begin{equation}
u_k^{\prime\prime}=\left\{-k^2+{1\over\eta^2} \left[\nu_S^2(\eta)-{1\over
4}\right]\right\}u_k,
\label{unu}
\end{equation}
where $\nu_S^2=(z''/z)\eta^2 + 1/4$ and the turning point is at
$k^2=\nu_S^2({\bar\eta}_S)/{\bar\eta}_S^2$. Note that the turning point,
${\bar{\eta}}_S$, is a given function of $k$. 

In an exactly analogous fashion, the equation for the tensor modes (\ref{modev})
is written in the form
\begin{equation}
v_k^{\prime\prime}=\left\{-k^2+{1\over\eta^2} \left[\nu_T^2(\eta)-{1\over
4}\right]\right\}v_k,
\end{equation}
where $\nu_T^2=(a''/a)\eta^2 + 1/4$ and the turning point is at
$k^2=\nu_T^2({\bar\eta}_T)/{{\bar\eta}_T}^2$. 

Unlike the approach of matching solutions through regions I, II, and
III, the idea of a uniform approximation is to provide a single
approximating solution which converges uniformly in all three regions
with a global, finite error bound. The normalization is determined
once and for all by simply matching to the exact solution as
$k\rightarrow \infty$.

To continue, we follow Olver in defining a new independent variable
$\xi$ and a new dependent variable $U$, given by~\cite{Olver}
\begin{equation}
\xi\left({d\xi\over d\eta}\right)^2=g(\eta),~~u=\left({d\xi\over
d\eta}\right)^{-1/2}U. 
\end{equation}
In terms of the new variables, Eqn.~(\ref{general}) becomes 
\begin{equation}
\label{dwequ}
{d^2U\over d^2\xi}=\left[b^2\xi+\psi(\xi)\right]U,
\end{equation}
where
\begin{eqnarray}\label{psixi}
\psi(\xi)&=&\left[4g(\eta)g''(\eta)-5g'^2(\eta)\right]
\frac{\xi}{16g^3(\eta)}\nonumber\\
&&+\frac{\xi q(\eta)}{g(\eta)}+\frac{5}{16\xi^2},\\
\frac 2 3 \xi^{3/2}&=&
-\int_{\eta}^{\bar\eta}\sqrt{g(\eta)}d\eta,~~~\eta \ge \bar\eta\\
\frac 2 3\left(-\xi\right)^{3/2}&=&
\int_{\eta}^{\bar\eta}\sqrt{- g(\eta)}d\eta,~~~\eta\le\bar\eta.
\end{eqnarray}
Now imagine neglecting $\psi(\xi)$ as a first approximation; then the
solution to the differential equation is given immediately in terms of
Airy functions. In the next order where $\psi(\xi)$ is no longer
neglected, the derivation of the solution $u$ becomes more involved.
Fortunately, in Ref.~\cite{Olver} the general solution for
Eqn.~(\ref{general}) in the uniform approximation to all orders is
derived with error bounds to be
\begin{eqnarray}\label{gensol1}
u_{2n+1}^{(1)}(b,\xi)&=&\left[\frac{g(\eta)}{\xi}\right]^{-1/4}
\left[{\rm Ai}(b^{2/3}\xi)\sum_{s=0}^n\frac{A_s(\xi)}{b^{2s}}
\right.\\
&&\left.+\frac{{\rm Ai}'(b^{2/3}\xi)}{b^{4/3}}
\sum_{s=0}^{n-1}\frac{B_s(\xi)}{b^{2s}}+\epsilon_{2n+1}^{(1)}\right],
\nonumber\\
\label{gensol2}
u_{2n+1}^{(2)}(b,\xi)&=&\left[\frac{g(\eta)}{\xi}\right]^{-1/4}
\left[{\rm Bi}(b^{2/3}\xi)\sum_{s=0}^n\frac{A_s(\xi)}{b^{2s}}
\right.\\
&&\left.+\frac{{\rm Bi}'(b^{2/3}\xi)}{b^{4/3}}
\sum_{s=0}^{n-1}\frac{B_s(\xi)}{b^{2s}}+\epsilon_{2n+1}^{(2)}\right],
\nonumber
\end{eqnarray}    
with coefficients defined by an iterative procedure,
\begin{eqnarray}
\hspace{-3mm}A_0(\xi)&=&1~~~~{\rm w.l.o.g.}, \\
\hspace{-3mm}B_s(\xi)&=&\frac{\pm
1}{2\sqrt{\pm\xi}}\int_0^\xi\left[\psi(v)A_s(v)-A_s''(v)\right] 
\frac{dv}{\sqrt{\pm v}},\label{Bs}\\
\label{Asp1}
\hspace{-3mm}A_{s+1}(\xi)&=&-\frac{1}{2}B'_s(\xi)+\frac 1
2\int\psi(\xi)B_s(\xi)d\xi,
\end{eqnarray}
where the upper signs are to be taken on the right of the turning
point $\bar\eta$ and the lower signs on the left of the turning point.
The error terms $\epsilon_{2n+1}^{(1)}$ and $\epsilon_{2n+1}^{(2)}$
are discussed in Appendix~\ref{error}. In the next and following
Sections, we will calculate the leading and next-to-leading order
solution for Eqn.~(\ref{unu}) with explicit error bounds and derive
the corresponding power spectra and spectral indices.

\section{Results at Leading Order}
\label{lead}

We now turn to the specific form of the approximating solutions at
leading order.  Taking $n=0$ in Eqns.~(\ref{gensol1}) and
(\ref{gensol2}) we find a solution for $u_k(\eta)$ containing a part
valid to the left of the turning point ($\eta\le\bar\eta$) and a part
valid to the right of the turning point ($\eta\ge\bar\eta$).  The
unnormalized solutions are
\begin{eqnarray}
u^{(1)}_{k,\lgtr}(\eta)&=&u^{(1)}_{k,1,\lgtr}(\eta)\left[
1+\epsilon_{k,1,\lgtr}^{(1)}(\eta)\right],
\label{uniformapp1}\\
u^{(2)}_{k,\lgtr}(\eta)&=&u^{(2)}_{k,1,\lgtr}(\eta)
\left[1+\epsilon_{k,1,\lgtr}^{(2)}(\eta)\right],
\label{uniformapp}
\end{eqnarray}
with
\begin{eqnarray}
u^{(1)}_{k,1,{\scriptscriptstyle\lessgtr}}(\eta)&=&
\left[f_{\scriptscriptstyle{\lessgtr}}(k,\eta)/
g_S(k,\eta)\right]^{1/4}{\rm Ai}[f_{\lgtr}(k,\eta)]\nonumber,\\
u^{(2)}_{k,1,\lgtr}(\eta)&=&\left[f_{\lgtr}(k,\eta)/
g_S(k,\eta)\right]^{1/4}
{\rm Bi}[f_{\lgtr}(k,\eta)],
\label{fsmgr}\\
f_{\lgtr}(k,\eta)&=&\mp\left\{\pm\frac 3
2\int_\eta^{\bar\eta_S}d \eta'\left[\mp 
g_S(k,\eta^{\prime})\right]^{1/2}\right\}^{2/3}\hspace{-3mm},\\
\label{gs}
g_S(k,\eta)&=&\frac{\nu_S^2(\eta)}{\eta^2}-k^2,\\
\label{eps1}
|\epsilon_{k,1,\lgtr}^{(1)}(\eta)|&\le&
\frac{1}{\lambda}\frac{M(f_{\lgtr})}{E(f_{\lgtr}){\rm Ai}(f_{\lgtr})}
\left\{e^{\lambda{\cal V}_{\eta,\alpha}({\cal E})}-1\right\},\\
\label{eps2}
|\epsilon_{k,1,\lgtr}^{(2)}(\eta)|&\le&
\frac{1}{\lambda}\frac{E(f_{\lgtr})M(f_{\lgtr})}{{\rm Bi}(f_{\lgtr})}
\left\{e^{\lambda{\cal V}_{\beta,\eta}({\cal E})}-1\right\},
\end{eqnarray}
where the lower index $1$ reflects the order of the approximation, the
functions with index $<$ are taken on the left of the turning point,
and those with the index $>$ are to be taken on the right of the turning
point.  $M(\eta)$, $N(\eta)$, ${\cal V}_{\alpha,\beta}$, and
$\lambda$ are defined in Appendix \ref{error}, and the error control
function ${\cal E}(\eta)$ is given by
\begin{eqnarray}
\label{ecf}    
{\cal E}(\eta)&=&-{1\over 4}\int
d\eta\left\{g_S^{-3/2}\left[g_S''-{5\over 
4}\frac{(g_S')^2}{g_S}-{g_S\over\eta^2}\right]\right\}\nonumber\\
&&\pm{5\over 24|f_{\lgtr}|^{3/2}}.
\end{eqnarray}
Inserting the explicit expression for $g_S(k,\eta)$, Eqn.~(\ref{gs}), and integrating
by parts leads to
\begin{eqnarray}
{\cal E}(\eta)&=&\frac{\nu_S(\nu_S'\eta-\nu_S)}{(\nu_S^2-k^2\eta^2)^{3/2}}
\pm{5\over 24|f_{\lgtr}|^{3/2}}\\
&&+\int\frac{d\eta}{4\eta\sqrt{\nu_S^2-k^2\eta^2}}\left\{
1-\left[\frac{\nu_S(\nu_S'\eta-\nu_S)}{\nu_S^2-k^2\eta^2}\right]^2\right\}.
\nonumber
\end{eqnarray}

The errors terms $\epsilon_{k,1,\lgtr}^{(1,2)}(\eta)$ in
Eqns.~(\ref{uniformapp1}) and (\ref{uniformapp}) encapsulate the
contributions to $u_{k,1,\lgtr}^{(1,2)}(\eta)$ beyond leading
order. The general solution for $u_k(\eta)$ is a linear combination of
the two fundamental solutions $u_k^{(1)}(\eta)$ and $u_k^{(2)}(\eta)$,
viz.,
\begin{equation}
u_k(\eta)=Au_k^{(1)}(\eta) + Bu_k^{(2)}(\eta), 
\end{equation}
independent of the order of the approximation.
In order to fix the constants $A$ and $B$ we have to construct a
linear combination of $u^{(1)}_k(\eta)$ and $u^{(2)}_k(\eta)$ such that
the result has the form $u_k(\eta)=e^{-ik\eta}/\sqrt{2k}$ in the
limit $k\rightarrow\infty$. In this limit, the domain of interest is
region I, far to the left of the turning point. In this case, for
well-behaved $\nu_S$, $f_<(k,\eta)$ is large and negative and we can
employ the asymptotic forms
\begin{eqnarray}
{\rm Ai}(-x)&=&\frac{1}{\pi^{1/2}x^{1/4}} \cos\left(\frac 2 3
x^{3/2}\right)-\frac{\pi}{4},\nonumber\\
{\rm Bi}(-x)&=&-\frac{1}{\pi^{1/2}x^{1/4}} \sin\left(\frac 2 3
x^{3/2}\right)-\frac{\pi}{4}.
\label{kbig}
\end{eqnarray}
Making the choices,
\begin{equation}
A=\sqrt{\frac \pi 2}e^{i\frac\pi 4},~~B=-i\sqrt{\frac \pi
2}e^{i\frac\pi 4}, 
\end{equation}
we find
\begin{equation}
u_{k,1,<}(\eta)=
\lim_{-k\eta\rightarrow \infty}{C\over \sqrt{2k}}~\exp
\left\{i\frac 3 2  \left[f_<(k,\eta)\right]^{3/2}\right\},
\label{bigkint}
\end{equation}
which is the required adiabatic form of the solution at short
wavelengths and as $\eta\rightarrow-\infty$~\cite{fulling}. $C$ is a
constant phase factor which is irrelevant when computing the
power spectrum.

The $\eta\rightarrow 0^-$ limit defines the region of interest for
calculating power spectra and the associated spectral indices.  In
this region, the $1/\eta^2$ pole dominates the behavior of the
solutions and the Airy solution goes over to the LG solution.  The LG
form of the solution is more tractable than the Airy form, leading to
simple expressions for the spectral indices.  We now demonstrate how
the Airy solution for small $\eta$ approaches the LG solution.  The
linear combination of Eqns.~(\ref{uniformapp1}) and (\ref{uniformapp})
in first order with the appropriate normalization is given by
\begin{equation}
u_{k,1,\lgtr}(\eta)=\sqrt{\frac{\pi}{2}}Cf_{\lgtr}^{1/4}(k,\eta)
g_S^{-1/4}(k,\eta)\left[{\rm Ai}(f_{\lgtr})-i{\rm Bi}(f_{\lgtr})\right],
\end{equation}
with the error bound
\begin{eqnarray}
|\epsilon_{k,1,\lgtr}(\eta)|&\le&
\frac{\sqrt{2}}{\lambda}
\left\{\right.
\left[\exp (\lambda{\cal V}_{\eta,\alpha}({\cal
E}))-1\right]\nonumber\\
&&+\left.\left[\exp (\lambda{\cal V}_{\beta,\eta}({\cal E}))-1\right]
\right\}\nonumber
\end{eqnarray}
derived from Eqns.~(\ref{eps1}), (\ref{eps2}), (\ref{Mx}), and
(\ref{Ex}). (The explicit form of the variation of ${\cal E}$ will be
discussed below.) For small $\eta$ we are on the right of the turning
point; the argument of the Airy functions, i.e., $f_>(k,\eta)$, becomes
large and the Airy functions can be approximated by
\begin{eqnarray}
{\rm Ai}(x)&=&\frac{1}{2\sqrt{\pi}}~x^{-1/4} \exp\left(-\frac{2}{3}
x^{2/3}\right),\\
\label{bilargex}
{\rm Bi}(x)&=&\frac{1}{\sqrt{\pi}}~x^{-1/4}
\exp\left(\frac{2}{3} x^{2/3}\right),
\end{eqnarray}
which leads to
\begin{eqnarray}
u_{k,1,>}(\eta)&=&{C\over\sqrt{2}}
g_S^{-1/4}(k,\eta) \left[{\frac 1 2}
\exp\left\{-\frac 2 3 \left[f_>(k,\eta)\right]^{3/2}
\right\}
\right.\nonumber\\
&&\left.-i \exp\left\{
\frac 2 3\left[f_>(k,\eta)\right]^{3/2}
\right\}
\right].
\end{eqnarray}
For computing the power spectra in the $k\eta\rightarrow 0^-$ limit, only
the growing part of the solution is relevant:
\begin{equation}
u_{k,1,>}(\eta)=
\lim_{k\eta\rightarrow 0^-}
-i C\sqrt{{-\eta\over 2\nu_S(\eta)}}
\exp\left\{
\frac 2 3\left[f_>(k,\eta)\right]^{3/2}
\right\}.\label{lguk}
\end{equation}

Once the approximate solutions to Eqns.~(\ref{mode}) and (\ref{modev})
have been found in the manner described above, the relevant power
spectra can easily be computed. The definition of the scalar power
spectrum for $\zeta$, where it is understood that all time-dependent
quantities are to be evaluated in the limit $\eta\rightarrow 0^-$, is
\begin{eqnarray}
P_S(k)&=&
\lim_{k\eta\rightarrow 0^-}
\frac{k^3}{2\pi^2}\left|\frac{u_k(\eta)}{z(\eta)}\right|^2\nonumber\\
&=&
\lim_{k\eta\rightarrow 0^-}
\frac{k^3}{2\pi^2}\left|\frac{u_{k,1,>}(\eta)}{z(\eta)}\right|\left|1+
\epsilon_{k,1,>}(\eta)\right|^2\nonumber\\
&=& 
\lim_{k\eta\rightarrow 0^-}
P_{1,S}(k)\left[
1+\epsilon_{k,1,S}^P(\eta)\right],
\label{Psp}
\end{eqnarray}
with
\begin{equation}
\label{scalerr}
\epsilon_{k,1,S}^P = 2\epsilon_{k,1,>},
\end{equation}
$P_{1,S}(k)$ denoting the power spectrum for the scalar perturbations in
the leading order approximation and dropping a second-order term in
$\epsilon_{k,1,>}$. Substituting the LG expression for $u_k$ from
Eqn.~(\ref{lguk}), we have 
\begin{equation}\label{PSP1}
P_{1,S}(k)=\lim_{k\eta\rightarrow 0^-}
\frac{k^3}{4\pi^2}\frac{1}{|z(\eta)|^2}
\frac{-\eta}{\nu_S(\eta)}
\exp\left\{
\frac 4 3 \left[f_>(k,\eta)\right]^{3/2}
\right\},
\end{equation}
with the error bound for the power spectrum given in Eqn.~(\ref{scalerr}).

We now discuss the variation of $\mathcal{E}$.  (See
Appendix~\ref{error} for a short discussion on the variation of a
function in general.) We are interested in the error bound of
$u_{k,1}$ over the full domain of interest $-\infty < \eta < 0^-$,
which implies $\beta=-\infty$ and $\eta\rightarrow 0^-$. In the general
case, 
\begin{equation}
{\cal V}_{-\infty,\eta}({\cal E})=\sum |{\cal E}(\alpha)-{\cal E}(\beta)|,
\end{equation}
where the sum is over all individual monotonic subintervals
$(\alpha,\beta) \subseteq (-\infty,\eta)$ of ${\cal E}$.  In the special case of monotonic
${\cal E}$ over the full range of $\eta$ the answer can be given in a
simplified form: By inserting the definition of $g_S(k,\eta)$ from
Eqn.~(\ref{gs}) into Eqn.~(\ref{ecf}) and integrating
by parts we find for the variation of the error control function:
\begin{eqnarray}
{\cal V}_{-\infty,\eta}({\cal E})&=&\left|-\frac{1}{2\nu_S}
-\frac 1 4\int\frac{d\eta}{\eta\sqrt{\nu_S^2-k^2\eta^2}}\right.\nonumber\\
&&\hspace{0.2cm}\left.\times\left\{
1-\left[\frac{\nu_S(\nu_S'\eta-\nu_S)}{\nu_S^2-k^2\eta^2}\right]^2
\right\}\right|,
\label{scalvar}  
\end{eqnarray}
where it is understood that $\eta$ has to be taken in the limit
$\eta\rightarrow 0^-$. 
With this expression the error bound for the power spectrum given in
Eqn.~(\ref{scalerr}) is completely determined.

The calculation for the tensor power spectrum follows along the same
lines, yielding
\begin{equation}\label{PTP1}
P_{1,T}(k)=
\lim_{k\eta\rightarrow 0^-}
\frac{k^3}{4\pi^2}\frac{1}{|a(\eta)|^2}
\frac{-\eta}{\nu_T(\eta)}
\exp\left\{
\frac 4 3\left[\tilde{f}_>(k,\eta)\right]^{3/2}
\right\},
\end{equation}
with the error being controlled exactly in the same manner as in
Eqns.~(\ref{scalerr}) and (\ref{scalvar}) with the subscripts
$S\rightarrow T$ and $\tilde f (k,\eta)$ indicating that $g_S(k,\eta)$
in Eqn.~(\ref{fsmgr}) has to be replaced by $g_T(k,\eta)$.


Before proceeding to the computation of the spectral indices,
we make a few observations on the nature of the error term
above. In previous work \cite{hhjm} we have shown that the
error term in the case of constant \(\nu\) introduces
a \(k\)-independent amplitude correction for the power spectrum
which does not affect the spectral index; therefore the
uniform approximation for the spectral index is exact
in that case. In a later section we explicitly
discuss this calculation.
It turns out to be possible to utilize this result in a
general way, and we turn to this now.

Suppose that we split the effective potential into two
terms, writing
\begin{equation}
  \nu^2(\eta) - \frac{1}{4} = \nu^2(0) - \frac{1}{4} + \nu^2(\eta) - \nu^2(0).
\end{equation}
Further we define a corresponding splitting of the error term,
\(\epsilon_{k,1,>} = \epsilon_0 + \tilde\epsilon\), where
\(\epsilon_0\) is the error term for the case of constant
\(\nu(\eta)\), the constant being given by \(\nu(0)\). Following
Olver, the total error term and the error term \(\epsilon_0\) each
satisfies an integral equation \cite{Olver}; using these we easily
derive an integral equation for \(\tilde\epsilon\) with a somewhat
different inhomogeneity.  By construction, this inhomogeneity is
reduced in size compared to the inhomogeneity which appears in the
integral equation for the full error term \(\epsilon\).  Using the
fact that the error term for constant \(\nu(\eta)\) is
\(k\)-independent, in this way we demonstrate that the full error term
can be written as a sum of two terms: \(\epsilon_0\) which is
\(k\)-independent and \(\tilde\epsilon\) which has the property that
it vanishes identically for constant \(\nu(\eta)\) and satisfies an
integral equation with a reduced inhomogeneity. Explicit calculation,
applying the theorem of Olver in the slightly generalized context,
gives a relative error bound of the form
\begin{equation}
  \left|{\tilde\epsilon(\eta)}\right|
     \le \frac{1}{\mathrm{Bi}(f_{\lgtr})} E(f_{\lgtr}) M(f_{\lgtr}) 
          \Phi(f_{\lgtr})
         \exp[\lambda \mathcal{V}_{\alpha,\eta}(\mathcal{E})],
\end{equation}
where \(E(f_{\lgtr})\) and \(M(f_{\lgtr})\) are the comparison
functions introduced in the appendix and used in the discussion above,
and where \(\Phi(f_{\lgtr}) = \mathcal{V}_{\alpha,\eta}(\mathcal{E} -
\mathcal{E}_0)\); here \(\mathcal{E}\) is the full error control
function and \(\mathcal{E}_0\) is the error control function for the
constant \(\nu(\eta)\) approximant.

This splitting therefore resums the error contributions which arise
purely from the constant part of \(\nu(\eta)\) and separates them into
an explicit \(k\)-independent additive contribution to the full error,
leaving a term which is quantitatively smaller and vanishes in the
case of constant \(\nu(\eta)\). This shows why we expect the uniform
approximation to the spectral index to be a much better approximation
than a direct application of the Olver theory would indicate.


Next we discuss the evaluation of the spectral indices.  The
generalized spectral index for scalar perturbations can be obtained
from the power spectrum via
\begin{equation}
n_S(k)=1+\frac{d\ln P_S}{d\ln k}.
\label{defns}
\end{equation}
Differentiation of the power spectrum with respect to $k$ is
straightforward. It is important to note that the turning point
$\bar\eta_S$ is a function of $k$, since
$k=-\nu_S(\bar\eta_S)/|\bar\eta_S|$ where $\nu_S(\bar\eta_S)$ is the
value of $\nu_S(\eta)$ at the turning point $\eta=\bar\eta_S$. Using
this relation, one finds
\begin{equation}
n_{1,S}(k)=4-2k^2
\lim_{k\eta\rightarrow 0^-}\int_{\bar\eta_S}^\eta\frac{d\eta'}
{\sqrt{g_S(k,\eta')}}.\label{nsint1}
\end{equation}

Following from the discussion above, the error in the spectral index
arises only from the $k$-dependent part of the error in the power
spectrum, which vanishes in the case of constant $\nu_S$. Thus the
error in the spectral index is sensitive only to the time variation of
$\nu_S$. To estimate this error, the spectral index as written in
Eqn.~(\ref{defns}) can be expressed via the leading order power
spectrum in the following form: 
\begin{eqnarray}
n_S(k)&=&1+\frac{d \ln(P_{1,S}+\epsilon_{k,1,S}^P)}{d\ln k}\nonumber\\
&=&1+\frac{d\ln P_{1,S}}{d\ln k}
+k\frac{d\epsilon_{k,1,S}^P}{d k}\nonumber\\
&\equiv&n_{1,S}(k)+\epsilon_{k,1,S}^n,
\end{eqnarray}
with
\begin{equation}
\label{errorn}
\epsilon_{k,1,S}^n = {k}\frac{d\epsilon_{k,1,S}^P}{d k}
\end{equation}
Here we have neglected error terms of order $\epsilon^2$.  We can
estimate $|\epsilon_{k,1,S}^n|$ by using the next-order contribution
to the power spectrum (Section \ref{nexttolead}), leading to the final
result 
\begin{eqnarray}
\epsilon_{k,1,S}^n(\eta)&\approx& -\frac{k^2}{\sqrt{2}}\int\frac{\eta
d\eta}{(\nu_S^2-k^2\eta^2)^{3/2}}\\ 
&&\times
\left[1-5\nu_S^2
\left(\frac{\nu_S-\nu_S'\eta}{\nu_S^2-k^2\eta^2}\right)^2\right].
\nonumber
\end{eqnarray}    
Again, it is understood that $\eta$ has to be taken in the limit
$\eta\rightarrow 0^-$, thus the spectral index defined in
Eqn.~(\ref{nsint1}) has no time-dependence.  

The above analysis can be carried out for tensor perturbations in an
identical fashion, including the error estimation, with the
replacement $\nu_S\rightarrow\nu_T$.  The spectral index for
gravitational waves is given by 
\begin{equation} 
n_{1,T}(k)=3-2k^2
\lim_{k\eta\rightarrow 0^-}
\int_{\bar\eta_T}^\eta\frac{d\eta'} {\sqrt{g_T(k,\eta')}}.
\label{ngint}
\end{equation}

Approximate evaluation of the integrals in Eqns.~(\ref{nsint1}) and
(\ref{ngint}) can be performed to yield more familiar forms for the
spectral indices as shown in Section \ref{LOCAPP}.

\section{Next-to-Leading Order}
\label{nexttolead}
To proceed to the next order, we first write the unnormalized solution
for $u_k$, following directly from Eqns.~(\ref{gensol1}) and
(\ref{gensol2}): 
\begin{eqnarray}
u^{(1)}_{k,3\lgtr}(\eta)&=&\left[f_{\lgtr}(k,\eta)/
g_S(k,\eta)\right]^{1/4}\left\{{\rm Ai}[f_{\lgtr}(k,\eta)]
\left(A_0[f_{\lgtr}(k,\eta)]
\right.\right.\nonumber\\
&&\hspace{-0.5cm}+\left.\left.A_1[f_{\lgtr}(k,\eta)]\right)
+{\rm Ai}'[f_{\lgtr}(k,\eta)]B_0[f_{\lgtr}(k,\eta)]\right\},\\
u^{(2)}_{k,3\lgtr}(\eta)&=&\left[f_{\lgtr}^{1/4}(k,\eta)/
g_S(k,\eta)\right]^{1/4}\left\{{\rm 
Bi}[f_{\lgtr}(k,\eta)]
\left(A_0[f_{\lgtr}(k,\eta)]
\right.\right.\nonumber\\
&&\hspace{-0.5cm}+\left.\left.A_1[f_{\lgtr}(k,\eta)]\right)
+{\rm Bi}'[f_{\lgtr}(k,\eta)]B_0[f_{\lgtr}(k,\eta)]\right\},    
\end{eqnarray}    
with
\begin{eqnarray}
A_0[f_{\lgtr}(k,\eta)]&=&1,\\
B_0[f_{\lgtr}(k,\eta)]&=&
\frac{\pm 1}{2\sqrt{\pm f_{\lgtr}(k,\eta)}}
\int_0^{f_{\lgtr}}\frac{\psi(v)}{\sqrt{\pm v}} dv,\\
A_1[f_{\lgtr}(k,\eta)]&=&-\frac 1 2 
B_0'[f_{\lgtr}(k,\eta)]\\
&&+\frac 1 2\int \psi[f_{\lgtr}]
B_0[f_{\lgtr}(k,\eta)] d[f_{\lgtr}(k,\eta)].
\end{eqnarray}  
The error bounds in next-to-leading order are given by [derived from
Eqns.~(\ref{err1}) and (\ref{err2})]:
\begin{eqnarray}    
|\epsilon_{k,3,\lgtr}^{(1)}|&\le& 
2E^{-1}(f_{\lgtr})M(f_{\lgtr}){\cal W}_{f_{\lgtr},\beta},\\
|\epsilon_{k,3,\lgtr}^{(2)}|&\le& 
2E(f_{\lgtr})M(f_{\lgtr}){\cal W}_{\alpha,f_{\lgtr}},
\end{eqnarray}    
with
\begin{eqnarray}
{\cal W}_{f_{\lgtr},\beta}&=&
\exp\left\{2\lambda \mathcal{V}_{f_{\lgtr},\beta}(|f_{\lgtr}|^{1/2}B_0)
\right\}\mathcal{V}_{f_{\lgtr},\beta}(|f_{\lgtr}|^{1/2}B_1),\nonumber\\ 
\\
{\cal W}_{\alpha,f_{\lgtr}}&=&
\exp\left\{2\lambda \mathcal{V}_{\alpha,f_{\lgtr}}(|f_{\lgtr}|^{1/2}B_0)
\right\}\mathcal{V}_{\alpha,f_{\lgtr}}(|f_{\lgtr}|^{1/2}B_1),\nonumber\\ 
\label{calw}
\end{eqnarray}
and
\begin{eqnarray}
B_{1}(f_{\lgtr})&=&\frac{\pm 1}{2\sqrt{\pm f_{\lgtr}}}
\int_0^{f_{\lgtr}}\frac{dv}{\sqrt{\pm 
v}}[\psi(v)A_1(v)-A''_1(v)],
\nonumber \\ 
\\
A''_1(v)&=&-\frac{1}{2}B_0'''(v)+\frac{1}{2}[\psi'(v)B_0(v)+\psi(v)B'_0(v)], 
\nonumber \\ 
\end{eqnarray}
recursively derived from Eqns. (\ref{Bs}) and (\ref{Asp1}).

As in leading order, the general solution $u_k(\eta)$ is a
linear combination of $u_k^{(1)}(\eta)$ and $u_k^{(2)}(\eta)$.
Fortunately, we will not have to calculate the normalization again; in
the limit $\eta\rightarrow -\infty$ the Bunch-Davies vacuum is the
exact solution of the differential equation for $u_k(\eta)$, and, in
this limit, all corrections from the next-to-leading order terms are
subdominant and of no interest.

A further simplification follows from the fact that only the growing
solution, $u_k^{(2)}$, is relevant to determining the power
spectrum and spectral index and that we can restrict ourselves to the
solution for $u_k$ in the limit $k\eta\rightarrow 0^-$. Employing once
again the approximation for the Bi-function for large, positive
argument, Eqn.~(\ref{bilargex}), and in addition, the approximation
for its derivative,
\begin{equation}
{\rm Bi}'(x)=\frac{1}{\sqrt{\pi}}~x^{1/4}
\exp\left(\frac{2}{3} x^{2/3}\right),   
\end{equation}    
the normalized $u_k(\eta)$ in the relevant regime is
\begin{eqnarray}
u_{k,3,>}(\eta)&\stackrel{k\eta\rightarrow 0^-}{=}&-i C
\sqrt{-{\eta\over\pi \nu_S(\eta)}}
\exp
\left\{\frac 2 3 \left[f_>(k,\eta)\right]^{3/2}
\right\}
\nonumber \\
&&\hspace{-.5cm}\times \left\{
1-\frac 1 2 B'_0(f_{>})
+\sqrt{f_{{>}}(k,\eta)}B_0(f_{>})\right.\nonumber\\
&&\left.+\frac 1 2\int \psi(f_{>})
B_0(f_{>}) d(f_{>})
\right\},
\end{eqnarray}
where the error bound is given by
\begin{equation}
|\epsilon_{k,3,>}|\le 2 f_>^{-1/4}
\exp{\left\{\frac 2 3 \left[f_>(k,\eta)\right]^{3/2} 
\right\}}
{\cal W}_{\alpha,f_{\lgtr}}.
\end{equation}
Analyzing $B_0$ in detail shows that the derivative $B_0'$ and the 
integral over $\psi B_0$ are subdominant in the limit $k\eta 
\rightarrow 0^-$. Hence the only term leading to a correction of the 
power spectrum is $B_0$ itself. From the general expression
(\ref{Psp}), the power spectrum at next-to-leading order is
\begin{equation}
\label{PSP2}    
P_{2,S}(k)\stackrel{k\eta\rightarrow 0^-}{=}
P_{1,S}(k)|1+2\sqrt{f_>(k,\eta)}B_0[f_{{>}}(k,\eta)]|,
\end{equation}
with $P_{1,S}(k)$ as defined in Eqn.~(\ref{PSP1}) and
\begin{eqnarray}
\label{B0}
B_0(f_>)&=&\frac{1}{2\sqrt{f_>}}\int_0^{f_>}\frac{\psi(v)}{\sqrt{v}}dv,\\
\label{psiv}
\psi(v)&=&\frac{5}{16v^2}+\frac{v\left(4g_Sg''_S-5g_S^{\prime 2}
\right)}{16g^3_S}
-\frac{v}{4\eta^2 g_S}.\label{psi}
\end{eqnarray}
The first term in $B_0$ [after writing out $\psi(v)$ according to
Eqn.~(\ref{psiv})] can be integrated immediately. The contribution
from the lower integration limit, which appears divergent at a first
glance, cancels with contributions from the other terms in the
integral. This can be shown by expanding $\psi(v)$ around zero
and integrating explicitly.
The error bound can be calculated in the same way as in leading order.
We find
\begin{equation}
P_S(k)=\lim_{k\eta\rightarrow 0^-}
\frac{k^3}{2\pi^2}\left|\frac{u_{k,3,>}(\eta)}{z(\eta)}
\right|^2\left[1 + \epsilon_{k,3,S}^P(\eta)\right],
\end{equation}
with
\begin{equation}
\epsilon_{k,3,S}^P = \frac{2\epsilon_{k,3,>}}{u_{k,3,>}(\eta)}.
\end{equation}
The spectral index at this order, $n_{2,S}$, as  calculated from its
definition (\ref{defns}), is given by 
\begin{equation}
n_{2,S}(k)=n_{1,S}(k)+\lim_{k\eta\rightarrow 0^-} \frac{2k}{|1+2\sqrt{f_>}B_0|}
\frac{\partial (\sqrt{f_>}B_0)}{\partial k},
\end{equation}
where $n_{1,S}(k)$ is given by Eqn.~(\ref{nsint1}). Evaluating the
derivative of $B_0$ with respect to $k$ leads to the following
expression for the spectral index:
\begin{eqnarray}\label{nsint2}
n_{2,S}(k)&=&n_{1,S}(k)+\frac{2k^2}{2|1+2\sqrt{f_>}B_0|}\\
&&\times\left[
\int_{\bar\eta_S}^{\eta}\frac{d\eta'}{\sqrt{g_S}}\left(
\frac{g_S''}{2g_S^2}-\frac{15 g_S'^2}{g_S^3}-\frac{1}{4\eta^2g_S}\right)
\right.\nonumber\\
&&\left.-\int_0^{f_>}\frac{dv}{4v^2}\left(\psi(v)-\frac{15}{8v^2}\right)
\int_{\bar\eta_S}^{\eta}\frac{d\eta'}{\sqrt{g_S}}
\right].\nonumber
\end{eqnarray}
The error estimate for the spectral index can be obtained in a similar way
as in leading order, just as for the power spectrum. We find
\begin{equation}
n_S(k)=n_{2,S}(k)+\epsilon_{k,3,S}^n(\eta),
\end{equation}
with
\begin{equation}
\epsilon_{k,3,S}^n(\eta) = 2k\frac{d}{dk}
\frac{\epsilon_{k,3,>}(\eta)}{u_{k,3,>}(\eta)}.
\end{equation}
In the case of the errors in next-to-leading order we have to evaluate
${\cal W}_{\alpha,f_{\lgtr}}$ which is defined in Eqn.~(\ref{calw}).
This can be done in principle but the result is rather long and
complicated and we do not write it out here explicitly. In a
forthcoming paper~\cite{shetal} we will examine different inflation
models numerically and show the corresponding results for the
next-to-leading order error estimates.

Proceeding in the same way as for the scalar perturbations the power
spectrum and the spectral index for the tensor perturbations,
including error bounds, can be calculated. $P_{2,T}(k)$ can be
obtained from Eqns.~(\ref{PSP2})-(\ref{psi}) by replacing 
$P_{1,S}(k)$ with $P_{1,T}(k)$ on the r.h.s. of Eqn.~(\ref{PSP2}) and
replacing $g_S(k,\eta)$ and its derivatives in Eqn.~(\ref{psi}) by
$g_T(k,\eta)$ and its derivatives. The spectral index for the tensor
perturbations can easily be derived by replacing $n_{1,S}(k)$ and
$g_S(k,\eta)$ and its derivatives in Eqn.~(\ref{nsint2}) on the right
hand side by $n_{1,T}(k)$ and $g_T(k,\eta)$ and its derivatives.

\section{Local Approximations}
\label{LOCAPP}
\subsection{Uniform Approximation}

The solutions obtained so far from the uniform approximation are
nonlocal; for the sake of simplicity and in order to compare with
conventional slow-roll results, local expressions are desirable even
if some accuracy is sacrificed thereby. This requires making
additional approximations regarding the integrals in
Eqns.~(\ref{nsint1}) and (\ref{ngint}). The analysis is again
identical for the scalar and tensor cases so we address the scalar
case first. The integrand has a square-root singularity at the turning
point, i.e., at the lower integral limit. At the upper limit $\eta$
goes to zero and the integrand vanishes linearly, therefore, assuming
$\nu_S(\eta)$ is well-behaved, we expect the main contribution to the
integral to arise from the lower limit. Combined with the knowledge
that $\nu_S$ is slowly varying it is reasonable to expand $\nu_S$
around the turning point in a Taylor series. To second order in
derivatives $\nu_S(\eta)$ reads
\begin{equation}
\nu_S^2(\eta)\simeq
\bar\nu_S^2+2\bar\nu_S\bar\nu'_S
\left(\eta-\bar\eta_S\right)
+({\bar\nu_S}^{\prime 2}+\bar\nu_S''\bar\nu_S)(\eta-\bar\eta_S)^2, 
\label{nuexp}
\end{equation}
where the bar indicates that this quantity has to be evaluated at the
turning point. We can now solve the integral in Eqn.~(\ref{nsint1})
exactly and find for the scalar spectral index
\begin{eqnarray}
\label{nslocal}
n_S(k)&\simeq&4-2\bar\nu_S\left\{1-\frac{\bar\nu'_S}{\bar\nu_S}\bar\eta_S
\left(1-\frac\pi 2\right)\right.\\
&&\left.+\frac{\bar\eta^2_S}{2}\left[
\frac{\bar\nu'^{2}_S}{\bar\nu^2_S}(2-\pi)
+\frac{\bar\nu''_S}{\bar\nu_S}(1-\pi)\right]\right\}.\nonumber
\end{eqnarray}
This is a simplification of the leading order result in the uniform
approximation; at the per cent level of accuracy for the spectral
index expected at this order, we have verified that it is adequate to
keep terms up to second derivatives of $\nu_S(\eta)$, as in
Eqn.~(\ref{nuexp}).

For the tensor spectral index we find analogously
\begin{eqnarray}
\label{ntlocal}
n_T(k)&\simeq&3-2\bar\nu_T\left\{1-\frac{\bar\nu'_T}{\bar\nu_T}\bar\eta_T
\left(1-\frac\pi 2\right)\right.\\
&&\left.+\frac{\bar\eta^2_T}{2}\left[
\frac{\bar\nu'^{2}_T}{\bar\nu^2_T}(2-\pi)
+\frac{\bar\nu''_T}{\bar\nu_T}(1-\pi)\right]\right\}.\nonumber
\end{eqnarray}

\subsection{Slow-Roll and its Variants}
\label{slowroll}

As a prelude to the comparison of slow-roll results with those
obtained from the local approximation discussed above, we give a brief
overview of the slow-roll paradigm for calculating the power
spectrum. Returning to the matching problem discussed in Section II
(represented by Fig.~\ref{fig1}), a rough estimate of the power
spectrum may be obtained by extrapolating the high and low-frequency
solutions to an intermediate regime $-k\eta=1$ or $k=aH$
(`horizon-crossing' for the $k$-mode of interest) and equating them at
that point.  It is understood that $\left|A_k\right|$ is determined
from the matching condition at $k=aH$, i.e.,
$\left|A_k\right|=1/(z\sqrt{2k})$.  With this substitution, the
familiar leading-order result ~\cite{guthpi}
\begin{equation}
P_S(k)\sim\left.\left(H\over 2\pi\right)^2\left(H^2\over
\dot{\phi}^2\right)\right|_{k=aH} \label{zeroP} 
\end{equation} 
is obtained. This expression is useful in providing quick estimates
for the power spectrum but is clearly not very precise: the solutions
are being matched in a region where they were not meant to be applied.
The conventional `slow-roll' approach to proceeding further is to
improve the matching by providing a better intermediate solution in
the region $-k\eta\sim 1$ or $k\sim aH$ (region II of Fig.~\ref{fig1})
and then to match the short and long wavelength solutions against the
intermediate solution. 

To understand the situation more concretely, it is useful to express
$z^{\prime\prime}/z$ and $a^{\prime\prime}/a$ in the exact forms 
\begin{eqnarray}
{z^{\prime\prime}\over z}&=&2a^2H^2\left(1+\epsilon+{3\over 2}\delta_1
+2\epsilon\delta_1+\epsilon^2+\frac 1 2 \delta_2\right),
\label{zdpz} \\
{a^{\prime\prime}\over a}&=&2a^2H^2\left(1-{1\over
2}\epsilon\right), \label{adpa} 
\end{eqnarray}
where
\begin{equation}
\epsilon\equiv-{\dot{H}\over H^2}=\frac 1
2\left(\frac{\dot\phi}{H}\right)^2, 
~\delta_n\equiv \frac{1}{H^n\dot\phi}\frac{d^{n+1}\phi}{dt^{n+1}}.
\label{epsdelxi}
\end{equation}
Here we have followed the notation of Stewart and Gong~\cite{sg},
which is especially convenient for comparing results expanded order by
order in slow-roll parameters as given in the next section. (This
convention is slightly different from that used in our previous
study~\cite{hhjm} which followed the conventions of
Ref.~\cite{LLKCBA}.) 

The derivatives of the leading order slow-roll parameters
$\epsilon$ and $\delta_1$ are of second order in the slow-roll
parameters:
\begin{eqnarray}\label{delta2der}
\frac{\dot{\epsilon}}{H}&=&2\delta_1\epsilon+2\epsilon^2,\\
\label{delta2der2}
\frac{\dot{\delta}_1}{H}&=&\delta_2+\delta_1\epsilon-\delta_1^2.
\end{eqnarray}
In principle, the parameters $\epsilon$ and $\delta_n$ are functions
of time that can be determined by solving the Friedmann equations.  In
the slow-roll approximation, the values of $\epsilon$ and $\delta_n$
are assumed to be small, and one aims to solve Eqns.~(\ref{mode}) and
(\ref{modev}) in terms of expansions in these parameters. At leading
order, with $\epsilon$ and $\delta_1$ both $\ll 1$, $\delta_2$ being
already next order can be neglected:
$\delta_2\sim{\cal{O}}(\epsilon^2, \delta^2_1, \epsilon\delta_1)$,
then it follows immediately from Eqns.~(\ref{delta2der}) and
(\ref{delta2der2}) that, to leading order, the derivatives of
$\epsilon$ and $\delta_1$ are approximately zero, and therefore they
can be treated as constants. It is crucial to note that at higher
order in the slow-roll expansion, the assumption that the parameters
are approximately constant does {\em not} hold, and the analysis
becomes complicated~\cite{wms}.

With these results in hand, the leading order slow-roll versions of
Eqns.~(\ref{mode}) and (\ref{modev}) become
\begin{eqnarray}
u_k^{\prime\prime}+\left[k^2 -{1\over\eta^2}\left(\nu_S^2-{1\over
4}\right)\right]u_k&=&0, \label{leadsrs}\\
v_k^{\prime\prime}+\left[k^2 -{1\over\eta^2}\left(\nu_T^2-{1\over
4}\right)\right]v_k&=&0, \label{leadsrgw}
\end{eqnarray}
where the leading order $\nu_S$ and $\nu_T$ are given by 
\begin{equation}
\nu_S={3\over 2}+2\epsilon+\delta_1,~~\nu_T={3\over 2}+\epsilon.
\label{leadnu}
\end{equation}
These equations can be solved in terms of Hankel functions as
\begin{eqnarray}
u_k(\eta)=\frac{\sqrt{\pi}}{2}e^{i(\nu_S+\frac{1}{2})\frac{\pi}{2}}
(-\eta)^{\frac 1 2}{H_{\nu_S}}^{(1)}(-k\eta),\\    
v_k(\eta)=\frac{\sqrt{\pi}}{2}e^{i(\nu_T+\frac{1}{2})\frac{\pi}{2}}
(-\eta)^{\frac 1 2}{H_{\nu_T}}^{(1)}(-k\eta).
\end{eqnarray}
The Hankel function solutions can now be matched to the short and long
wavelength solutions as described in Ref.~\cite{wms} (see also
Ref.~\cite{ms}). To calculate the power spectrum we need in addition
an expression that connects the conformal time $\eta$ with the
slow-roll parameters. This can be obtained to any order by repeated
integration by parts~\cite{LLKCBA}: 
\begin{eqnarray}
\eta&\simeq&-{1\over 
aH}\left[1+\epsilon+3\epsilon^2+2\epsilon\delta_1+15\epsilon^3\right.
\nonumber\\
&&\left.+20\epsilon^2\delta_1+2\epsilon\delta_1^2+2\epsilon\delta_2
+ \cdots\right].
\label{etasr}
\end{eqnarray}
It follows from Eqn.~({\ref{etasr}) that at leading order,
$\eta\simeq-(1+\epsilon)/aH$. Using this result and the small argument
approximation for the Hankel functions, the first-order slow-roll
corrections to the scalar power spectrum (\ref{zeroP}), as originally
computed by Stewart and Lyth~\cite{sl}, read: 
\begin{equation}
P_{S}(k)\simeq[1-(2C+1)\epsilon+C\delta_1]\left.\left(H\over
2\pi\right)^2\left(H^2\over \dot{\phi}^2\right)\right|_{k=aH},
\label{leadP} 
\end{equation}
where $C=-2+\ln 2+\gamma\simeq-0.73$ and $\gamma$ is Euler's
constant. Errors inherent to the matching procedure and use of the
Hankel approximation are discussed in Refs.~\cite{wms} and~\cite{ms}.

As already explained above, straightforward extension of the slow-roll
approach to higher orders runs into serious
difficulties~\cite{wms,ms}. The primary obstacle is the fact that at
next-to-leading order, as is clear from Eqn.~(\ref{delta2der}), the
slow-roll parameters can no longer be treated as constants and the
Hankel solution no longer holds. Thus, to obtain more accurate
results, one has to abandon the original slow-roll expansion technique
of fixing a Hankel solution in the intermediate region and aiming to
obtain higher-order expressions for the Hankel index in terms of an
expansion in slow-roll parameters.

An alternative approach within the slow-roll methodology is the work
of Stewart and Gong~\cite{sg}. In this calculational scheme, explicit
matching of solutions is avoided by using a perturbative
approximation, allowing the inclusion of higher-order corrections (The
leading order solution is still the Bessel approximation of
Ref.~\cite{sl}.) To be more concrete, Stewart and Gong employ the
following approach: instead of replacing $z''/z$ in Eqn.~(\ref{mode})
by a constant divided by $\eta^2$ as in the original
slow-approximation, they choose the ansatz $z=\eta^{-1}f(\ln\eta)$.
This ansatz possesses, in addition to the term proportional to
$1/\eta^2$, a contribution which depends logarithmically on time. In
the differential equation for $u_k$ this additional term is then
treated as an inhomogeneity and the solution for $u_k$ can be found
using Green's function methods.  The final expression for $u_k$ is an
integral equation.  The integral in the expression for $u_k$ is then
expanded in slow-roll parameters.  By including a weak time dependence
in $z''/z$ this approach allows the extension of the slow-roll
approximation to higher orders without violating
Eqn.~(\ref{delta2der}).  Nevertheless, this approach also lacks an
error estimate. In a second paper~\cite{srimprov}, Stewart elaborates
on these results, though only at leading order, by choosing a general
expansion point for the slow-roll parameters (instead of $k=aH$) and
showing that expressing the slow-roll parameters in terms of potential
derivatives (see, e.g., Ref.~\cite{LPB} for a discussion of this
approach) can lead to incorrect results. An extension of this work
to one more order has been recently performed \cite{CGS}.

\subsection{Slow-Roll Redux}
\label{SRR}

We are now in a position to discuss our results in the context of the
slow-roll analysis of the previous subsection. In our case, obtaining
spectral indices from Eqns.~(\ref{nslocal}) and (\ref{ntlocal}) as a
function of $\epsilon$ and $\delta_n$ is straightforward: we simply
expand $\nu_S$, $\nu_T$, and their derivatives in terms of slow-roll
parameters.  A crucial point to keep in mind is the value of $k$ in
terms of which the results are stated. In the slow-roll expansion the
evaluation of the slow-roll parameters is traditionally given at
horizon crossing, $-k\eta\sim 1$, a somewhat arbitrary choice due to
the uncontrolled nature of the approximation. As shown in detail in
Ref.~\cite{srimprov} calculating the slow-roll parameters at a
convenient time close to horizon crossing leads to small finite
corrections in power spectra and spectral indices. In our case, the
natural expansion point is the turning point, therefore a truly direct
comparison with slow-roll results, such as those of Ref.~\cite{sg}, is
rather complicated.

We begin by considering the spectral index for the scalar
perturbations as given in Eqn.~(\ref{nslocal}). In order to write
$n_S$ in slow-roll parameters up to second order we expand
$\bar\nu_S$, $\bar\nu'_S \bar\eta_S$, and $\bar\nu''_S \bar\eta_S^2$
up to third order in slow-roll parameters. The contribution
$\bar{\nu_S'}^2\bar\eta_S^2/\bar\nu_S$ is already of fourth order and is
neglected. Using the expression for conformal time in slow-roll
parameters given in Eqn.~(\ref{etasr}), the relation between $\nu_S$
and $z''/z$,
\begin{equation}
\nu_S^2=\frac{z''}{z}\eta^2+\frac{1}{4},
\end{equation}
and the expression for $z''/z$ given in Eqn.~(\ref{zdpz}), we find for
the different contributions:
\begin{eqnarray}
\bar\nu_S&=&
{3\over 2}+2\bar\epsilon +\bar\delta_1 +{16\over 3} \bar\epsilon^2
+{14\over 3} \bar\epsilon\bar\delta_1 -{1\over 3}\bar\delta_1^2 
\nonumber\\
&&+{1\over 3} \bar\delta_2+{206\over 9} \bar\epsilon^3
+ {2 \over 9} \bar\delta^3_{1}
+ 4\bar\epsilon\bar\delta_1^2
+{287\over 9}\bar\epsilon^2\bar\delta_1\nonumber\\
&&+{26\over 9} \bar\epsilon\bar\delta_2
-{2\over 9 }\bar\delta_1\bar\delta_2
+{\cal O}(\bar\epsilon^4),\label{nubar}\\
{\bar\eta_S}{\bar\nu}'_S&=&-4\bar\epsilon-5\bar\epsilon\bar\delta_1
+\bar\delta_1^2-\bar\delta_2
-{121\over 3} \bar\epsilon^2\bar\delta_1
-\frac{76}{3}\bar\epsilon^3-3\bar\epsilon\bar\delta_1^2
\nonumber\\
&&-{19\over 3} \epsilon\bar\delta_2-\frac{2}{3}\bar\delta_1^3
+\bar\delta_1\bar\delta_2-{1\over 3}\bar\delta_3+{\cal
O}(\bar\epsilon^4),\\ 
\bar\eta^2_S\bar{\nu}''_S&=&
4\bar\epsilon^2+5\bar\epsilon\bar\delta_1-\bar\delta_1^2+\bar\delta_2
+\frac{124}{3}\bar\epsilon^3+\frac{214}{3}\bar\epsilon^2\bar\delta_1
-4\bar\delta_1\bar\delta_2
\nonumber\\
&&+\frac{40}{3}\bar\epsilon\bar\delta_2+\frac{8}{3}\bar\delta_1^3
+6\bar\epsilon\bar\delta_1^2
+\frac 4 3 \bar\delta_3+{\cal O}(\bar\epsilon^4).
\label{nubarpp}
\end{eqnarray}
As in earlier sections, the bar indicates that the slow-roll
parameters are to be calculated at the turning point, and ${\cal
O}(\bar\epsilon^4)$ represents all slow-roll terms of fourth order and
higher. Finally, inserting Eqns.~(\ref{nubar})-(\ref{nubarpp}) into
the local expression for the scalar spectral index (\ref{nslocal})
allows us to write the spectral index in terms of slow-roll
parameters: 
\begin{eqnarray}
n_S(k)&\simeq& 1-4\bar{\epsilon}-2\bar{\delta}_1-
8\bar{\epsilon}^2\left(\frac{17}{6}-\pi\right)\nonumber\\
&&-10\bar{\epsilon}\bar{\delta}_1\left(\frac{73}{30}-\pi\right)
+2(\bar{\delta}_1^2-\bar{\delta}_2)\left(\frac{11}{6}-\pi\right)\nonumber\\
&&-\frac{200}{3}\bar{\epsilon}^3\left(\frac{31}{15}-\pi\right)
-\frac{10}{3}\bar{\delta}_1^3\left(\frac{4}{3}-\pi\right)\nonumber\\
&&-\frac{335}{3}\bar{\epsilon}^2
\bar{\delta}_1\left(\frac{1942}{1005}-\pi\right)
-9\bar{\epsilon}\bar{\delta}_1^2\left(\frac{20}{9}-\pi\right)\nonumber\\
&&-\frac{59}{3}\bar{\epsilon}\bar{\delta}_2\left(\frac{286}{177}-\pi\right)
+5\bar{\delta}_1\bar{\delta}_2\left(\frac{58}{45}-\pi\right)\nonumber\\
&&-\frac{5}{3}\bar{\delta}_3\left(\frac{6}{5}-\pi\right).
\label{ns_slow}
\end{eqnarray}
Thus, in the end, starting from the nonlocal expression for the scalar
spectral index given by the uniform approximation
[Eqn.~(\ref{nsint1})], we have finally arrived at a local expression
for $n_S$ in terms of slow-roll parameters by employing two
expansions: First we expanded the integrand in the expression for the
spectral index in a derivative expansion in $\bar\nu_S$, to solve the
integral in Eqn.~(\ref{nsint1}).  Then we further expanded the result
in slow-roll parameters. However, the two expansions are not
independent; had we decided to stop the expansion in derivatives in
$\nu_S$ after the first term we would not have obtained the second
order slow-roll contributions from the expansion of the second
derivative of $\nu_S$. If one wants results quoted to some order in
slow-roll parameters, this requires going up to a finite order in
derivatives of $\bar\nu_S$; however, written in this way, it is also
clear that {\em a priori} it is not obvious which expansion is the
dominant one -- the expansion in derivatives of $\bar\nu_S$ or the
expansion in slow-roll parameters. In the absence of further
information regarding $\nu_S$ itself, the question cannot be answered
satisfactorily. This demonstrates one of the inherent difficulties of
deriving higher order expressions for the spectral index via Taylor
expansions without having a well-defined error bound.

The analogous result in Ref.~\cite{sg} agrees at leading order. The
forms of the higher order contributions are apparently different due
to the difference in evaluation points and the different
approximations employed. However, both results should be treated with
some caution: (i) Without an error control theory, it is not clear
that inclusion of higher order terms actually improves the accuracy of
the result (convergence is not guaranteed since the Taylor expansion
leads only to an asymptotic expansion) (ii) the evaluation point of
the slow-roll parameters leads to an uncertainty in the calculation --
if one really wants results accurate to the per cent level, this
uncertainty is important. In order to obtain results with high
accuracy, error controlled approximations appear to be necessary.

For completeness we also give the result for the tensor spectral index
expressed in slow-roll parameters. Since Stewart and Gong do not derive
an equivalent expression, we compare it instead to the quadratic-order
slow-roll result obtained originally by Stewart and
Lyth~\cite{sl}. For the tensor spectral index derived from a slow-roll
expansion of the local result Eqn.~(\ref{ntlocal}), we find
\begin{equation}
n_T(k)=-2\bar\epsilon-2\left(\frac{23}{4}-\pi\right)\bar\epsilon^2
+2\left(\frac{14}{3}-\pi\right)
\bar\epsilon\bar\delta_1,
\end{equation}
whereas Ref.~\cite{sl} obtains
\begin{equation}
n_T(k)=-2\epsilon-(3+c)\epsilon^2+(1+c)\epsilon\delta_1,
\end{equation}
with $c=0.08145$. As for the scalar spectral index, the leading order
contributions agree. 

\section{The Special Case of Constant $\nu$}
\label{special}

In order to demonstrate the accuracy of our approximation explicitly
we now investigate a special class of exactly solvable inflation
models, where $\nu_S$ and $\nu_T$ are constant. This class includes
models such as power-law inflation or inflation near a potential
maximum.  We restrict ourselves in this section to scalar
perturbations;  tensor perturbations can be treated in the same
way. The exact power spectrum for time-independent $\nu_S$ evaluated
from the general expression (\ref{Psp}) is
\begin{eqnarray}
P_S^{\rm ex}(k)&=&\frac{2^{2\nu_S-2}}{2\pi^3}
\Gamma^2(\nu_S)\left(\frac{H}{a\dot\phi}\right)^2
(-k\eta)^{1-2\nu_S}k^2\label{psex}\\
&=&\frac{2^{2\nu_S-2}}{\pi^2}
e^{-2\nu_S}\nu_S^{2\nu_S-1}\left(\frac{H}{a\dot\phi}\right)^2
(-k\eta)^{1-2\nu_S}k^2\nonumber\\
&&\times\left(1+\frac{1}{6\nu_S}+\frac{1}{72\nu_S^2}+\cdots\right),
\end{eqnarray}
where we have used Stirling's formula to replace the $\Gamma$
function.  [For a detailed derivation of Eqn. (\ref{psex}) see, e.g.,
Ref.~\cite{LLKCBA}.] The spectral index is easily found from
Eqn. (\ref{defns}): 
\begin{equation}
n_S=4-2\nu_S.
\end{equation}
The general expression for the power spectrum in leading and
next-to-leading order in the uniform approximation is given by
Eqn.~(\ref{PSP1}) and Eqn.~(\ref{PSP2}), respectively. For constant
$\nu_S$, the integrals which appear in these expressions can be solved
exactly, leading to the results:
\begin{eqnarray}
P_S^{(1)}(k)&=&\frac{2^{2\nu_S-2}}{\pi^2}
e^{-2\nu_S}\nu_S^{2\nu_S-1}\left(\frac{H}{a\dot\phi}\right)^2
(-k\eta)^{1-2\nu_S}k^2,
\nonumber\\
\label{PS1nuc}\\
P_S^{(2)}(k) &=&\frac{2^{2\nu_S-2}}{\pi^2}
e^{-2\nu_S}\nu_S^{2\nu_S-1}\left(\frac{H}{a\dot\phi}\right)^2
(-k\eta)^{1-2\nu_S}k^2\nonumber\\
&&\times\left(1+\frac{1}{6\nu_S}\right).
\label{PS2nuc}
\end{eqnarray}
Comparison of these results with the exact power spectrum reveals a nice
feature of our approximation for this special case: Improving our
approximation order by order leads to matching corrections to the
$\Gamma$-function in powers of inverse $\nu$. 

A more rigorous analysis of the errors in the uniform approximation
along the lines explained in Sections~\ref{lead} and \ref{nexttolead}
shows that the leading order solution is bounded by the absolute value
of the relative error
\begin{equation}
|\epsilon_{1,2}|\leq \sqrt{2}\left({1\over 6\nu_S}+{\lambda\over
 72\nu^2_S}+\cdots\right),
\end{equation}
where $\lambda\simeq 1.04$. The error in the power spectrum given in
Eqn.~(\ref{PS1nuc}) falls comfortably within the bound. 

Similarly, we can solve the integrals in the expressions for the 
spectral indices in leading and next-to-leading order, Eqns. 
(\ref{nsint1}) and (\ref{nsint2}), exactly. We find that the spectral
index in leading order is already exact:
\begin{equation}
n_S^{(1)}=n_S^{(2)}=4-2\nu_S.    
\end{equation}
This is an important result demonstrating the high accuracy of the
uniform approximation already at leading order. There are no 
higher order corrections, to be expected since the
corrections to the power spectrum in second order, Eqn.
(\ref{PS2nuc}), are $k$-independent. We can also evaluate the error
bound for the spectral index from Eqn.~(\ref{errorn}) for constant
$\nu$: Consistent with obtaining the exact value of $n_S$, we find
that in the small $k\eta$ limit, the error vanishes.

In order to demonstrate quantitatively the high accuracy of the
uniform approximation, we now consider the case of power-law inflation,
where $\nu$ is constant. In this case the scale factor evolves
as a power-law in time, $a\propto t^p$, hence the Hubble parameter
\begin{equation}
H(t)=pt^{-1}.
\end{equation}    
The power $p$ and $\nu_S$ are connected via
\begin{equation}
\nu_S=\frac 3 2+\frac{1}{p-1}.    
\end{equation}    
The conformal time is given by the exact expression
\begin{equation}
\eta=-\frac{1}{aH}{1\over (1/p-1)}.
\end{equation}    

Slow-roll is known to be inaccurate for small values of $p$, we
therefore pick $p=2$ as a test case -- equivalent to
$\nu_S=5/2$. Expressions for the exact power spectrum and the ones
obtained with the uniform approximation are already given above; we
now compare these with the conventional slow-roll results.
Traditionally results for slow-roll inflation are calculated at
$k=aH$. Following this choice, the result of Stewart and
Gong~\cite{sg} for power-law inflation is
\begin{eqnarray}
P_S^{SR}(k)&=&\frac{H^4}{4\pi^2\dot \phi^2}\left[1-{2\over p}(c+1)
\right.\nonumber\\
&&+\left.{2\over p^2}\left(c^2+c-{5\over 2}+\frac{\pi^2}{4}
\right)\right].    
\end{eqnarray}    

Figs.~\ref{fig2} and \ref{fig3} show a comparison of the exact power
spectrum versus wave number, with leading and next-to-leading
approximations from the slow-roll and the uniform approximation.  The
solid black line in both Figures represents the exact solution.  In
Fig.~\ref{fig2} we compare the first order approximations to the exact
solution. The dotted line shows the slow-roll result, which has an
error of almost 30\%, while the dashed line shows the power spectrum
obtained from the uniform approximation with an error of roughly
7\%. Fig.~\ref{fig3} displays the next-to-leading order result.  The
slow-roll result (dotted line) has improved but is still inaccurate to
9\%, which is worse than the result from the uniform approximation at
leading order. The grey dashed line (almost indistinguishable from the
black line) shows the next-to-leading order result from the uniform
approximation, which has the remarkably small error of only
0.4\%. This result is very encouraging; in a forthcoming
paper~\cite{shetal} we will demonstrate the excellent performance of
the uniform approximation for models which can only be solved
numerically.

\begin{figure}[h]
\leavevmode\includegraphics[width=8.2cm]{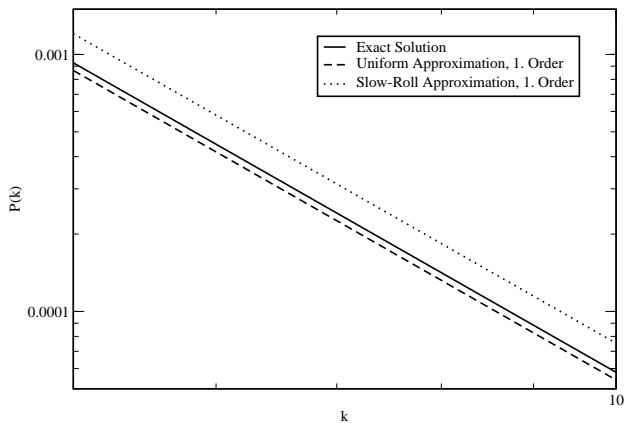}
\caption{Comparison of the exact power spectrum for power-law inflation with
the uniform approximation and the slow-roll approximation in leading order.} 
\label{fig2}
\end{figure}

\begin{figure}[h]
\leavevmode\includegraphics[width=8.2cm]{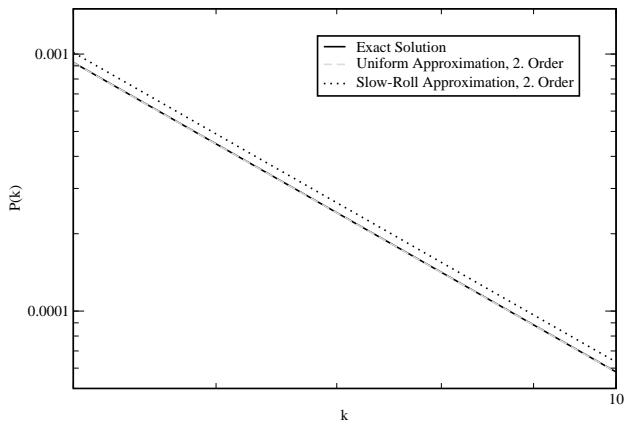}
\caption{Comparison of the exact power spectrum for power-law
inflation with the uniform approximation and the slow-roll
approximation in next-to-leading order. The uniform approximation
and the exact solution are indistinguishable in this plot.}
\label{fig3}
\end{figure}

In addition we can compare the spectral index for power law inflation.
As explained earlier, the uniform approximation leads to the exact
answer for the spectral index in the case of constant $\nu$. The
slow-roll answer for power law inflation up to second order in
slow-roll parameters is given by
\begin{equation}
n_S^{SR}=1-\frac{2}{p}-\frac{2}{p^2}.
\end{equation}
For the case $p=2$, the slow-roll approximation clearly does not lead
to a good answer, as the exact result is $n_S=-1$ while the slow-roll
answer is $n_S=-0.5$. Of course $p=2$ does not represent a realistic
model for inflation. At higher values of $p$ the slow-roll answer
improves dramatically, nevertheless, unlike the uniform approximation,
it is never exact. (The amplitude of the power spectrum in slow-roll
approximation also improves rapidly for larger values of $p$.) 

The remarkable feature of the uniform approximation is that the error
in the leading order result for the solution itself (and therefore for
the power spectrum) is always $\sim 8\% $, while the next-to-leading
order error is less than 1\%, independent of the choice of $p$. This
robustness of the error control is a key feature of the uniform
approximation.


\section{Comments on Bessel Function Approximants}
\label{besselcom}

Before concluding we would like to make another comparison to the
commonly used procedure involving Bessel functions as approximating
solutions for the fundamental wave equation. Recall that this
approximation arises by assuming that \(\nu(\eta)\) is ''locally''
constant, and it forms the basis for methods discussed previously in
the literature \cite{wms}.

From the standpoint of the Olver uniform theory, it turns out that
Bessel function approximants arise naturally when treating the case of
an ODE having a coefficient function with a pole of order two
\cite{Olver}. Since such a pole arises in the model equation at hand,
one might ask if this theory is not better suited to the computation
of the power spectrum, perhaps thereby making a connection to the
methods already known in the literature.

In fact, direct computation of the first order term in the Olver
theory with Bessel approximants shows that it is not possible to
obtain a reliable power spectrum in this way. The failure of the
method is essentially due to the failure of the matching procedure at
early times, \(\eta\rightarrow -\infty\). This failure arises because
the Bessel approximants, although uniformly controlled in \(\eta\),
lose too much of the more delicate \(k\)-dependence of the solution at
large \(\eta\), notably for small \(k\). Following Olver's discussion
of the relation between the Bessel approximants and the Born
approximation in quantum mechanics \cite{Olver}, another way to state
the problem is to note that the Born approximation will always fail
for sufficiently low energy.  In essence, Bessel approximants provide
a systematic way to calculate the power spectrum at \emph{high} \(k\),
whereas we need information at low \(k\).

The success of the Airy approximants is a result of a tradeoff. By
concentrating on the turning point at the expense of some early and
late time information, the Airy approximants provide a successful
bridge for the matching operation which is required. One of the costs
of this tradeoff is a non-vanishing error in the limit
\(\eta\rightarrow 0\), whereas the Bessel approximants have a
vanishing error in this limit. However, the error of the Bessel
approximants is not sufficiently controlled in the other limit,
\(\eta\rightarrow-\infty\), for small \(k\). Furthermore, we have
shown above how the bulk of the non-vanishing contribution to the
error for the Airy approximants gives rise to a \(k\)-independent
amplitude correction, which cannot effect the spectral index; this
significantly improves a priori expectations for the uniform
Airy-approximation.


\section{Conclusion}
\label{conclusion}

In this paper we have presented the uniform approximation as an
excellent technique for obtaining power spectra and spectral indices
from inflation models, making only a minimum number of dynamical
assumptions: the answers are stated in terms of elementary functions
and simple integrals which are easy to evaluate numerically. The
existence of calculable and robust error bounds is a crucial advantage
of the method.

In order to completely utilize this approach, the next step is a fast
numerical implementation, now in progress~\cite{shetal}. Combined with
a numerical code like CMBFAST~\cite{cmbfast} which translates
primordial fluctuations into the linear radiation and matter power
spectra, and with our now fully developed error analysis, such a
capability promises to be very useful for obtaining both forward
predictions from specific models as well as backward constraints from
observational data.

\section{Acknowledgements}

We thank Fabio Finelli and Ewan Stewart for helpful discussions. SH
and KH acknowledge the Aspen Center for Physics where part of this
work was completed.  AH gratefully acknowledges Los Alamos National
Laboratory for warm hospitality and thanks the Graduiertenkolleg
``Physics of Elementary Particles at Accelerators and in the
Universe'' for partial financial support. CM-P thanks the Nuffield
Foundation for partial financial support.

\begin{appendix}
\section{Gauge-invariant Perturbations}
\label{backg}

In order to fix notation and to introduce the main equation of
interest, we summarize results from cosmological perturbation theory
which are of direct interest for inflationary models. The basic
formalism is that of Bardeen~\cite{givb}; see Ref.~\cite{ivrevs2} for
a detailed discussion based on this formalism.

A homogeneous and isotropic Friedmann-Robertson-Walker (FRW) universe
is described by the metric
\begin{eqnarray}
  ds^2 &=& g_{a b} dx^a dx^b\nonumber\\
       &=& [-dt^2+ a^2(t)\gamma_{ij}(\vec x) dx^i dx^j]\nonumber\\
       &=& a^2(\eta)[-d\eta^2+ \gamma_{ij}(\vec x) d x^i dx^j ],
\label{bmet}
\end{eqnarray}
where \(\gamma_{ij}\) is the metric on homogeneous
and isotropic spatial sections, and
the conformal time is $\eta=\int^t dt'/a(t')$.
Let \(D_i\) be the covariant derivative on the spatial sections.
Latin letters denote spatial indices.
Considering only so-called scalar and tensor perturbations,
we write the perturbed metric in the form
\begin{equation}
  ds^2 = a(\eta)^2 \left[ds^2_S + ds^2_T\right],
\end{equation}
with the scalar and tensor perturbations
\begin{eqnarray}
  ds^2_S &=& -(1+2A)d\eta^2 + 2 D_i B d\eta dx^i \\
            && + \left[(1+2C)\gamma_{ij}
      + 2 \left(D_i D_j - \frac{1}{3}\gamma_{ij} \lapthree\right) E
                      \right] dx^i dx^j \nonumber\\
  ds^2_T &=& E_{ij} dx^i dx^j.
\label{pmet}
\end{eqnarray}
The tensor \(E_{ij}\) is transverse, symmetric, and traceless.
Gauge transformations of this perturbed metric are generated
by vector fields of the form \(\xi^a=(T,D^i L)\).

The two standard gauge invariant combinations which describe the scalar
perturbations of the metric are
\begin{eqnarray}
\Phi_A &=& A + (B' + h B) - (E'' + h E'),\\
\Phi_C &=& C - \frac{1}{3} \lapthree E + h (B - E').
\label{givs}
\end{eqnarray} 
Primes denote differentiation with respect to conformal time and
$h\equiv a'/a$. Note that in the longitudinal gauge
($L=-E$, $T=B-E'$), the perturbed metric becomes
\begin{equation}
  ds^2 = a^2(\eta)\left[-(1+2\Phi_A)d\eta^2 + (1+2\Phi_C)\gamma_{ij}dx^{i}dx^{j}\right].
\end{equation}
This gauge yields a direct physical interpretation for
\(\Phi_A\) and \(\Phi_C\).
The tensor perturbations described by \(E_{ij}\)
are themselves gauge-invariant.

In order to complete the setup for application of the Einstein
equations, we must introduce matter perturbations. The energy momentum
tensor for an FRW cosmology is that of a homogeneous and isotropic perfect fluid
\begin{equation}
  T_{ab} = \varrho(\eta) u_a u_b + p(\eta) (u_a u_b + g_{ab}),
\end{equation}
where the energy density $\varrho$ and the pressure $p$ are functions
only of time and $u_a(\eta)$ is the fluid four-velocity, which is comoving
with the gravitational background, $u_a = -a(\eta) (d\eta)_a$. The most
general form for the perturbation of the energy momentum tensor is
\begin{equation}
\delta T_{ab} = \delta \varrho u_a u_b + 2 a q_{(a} u_{b)} 
      + \delta p (u_a u_b + g_{ab}) + p a^2 \Pi_{ab},
\end{equation}
where $q_a$ is the perturbed velocity field of the fluid, and the
tensor $\Pi_{ab}$ represents the anisotropies of the
perturbations. The perturbations must satisfy the constraints
\begin{equation}
  g^{ab}q_a u_b=0,\quad g^{ab}\Pi_{ab}=0,\quad \Pi^{ab} u_b=0.
\end{equation}

Consider now the case of perturbations with vanishing
anisotropic stress, \(\Pi_{ab} = 0\). One consequence of the
Einstein equations in this case is that \(\Phi \equiv \Phi_A = \Phi_C\).
Further assume that the perturbations are adiabatic; then the Einstein
equations can be combined to give \cite{ivrevs2}
\begin{equation}
  \Phi'' + 3h(1 + c_s^2)\Phi' - c_s^2 \lapthree \Phi
   + (2h' + (1 + 3c_s^2)(h^2 - \mathcal{K})) \Phi = 0,
\end{equation}
where \(c_s(\eta)^2 \equiv p'/\varrho'\) has the interpretation of
the adiabatic sound speed squared, and \(\mathcal{K}\) is the
curvature of spatial sections. It turns out that if one defines
\(w\) by  \(\Phi \equiv 4\pi G (\varrho + p)^{1/2} w\), then
\begin{equation}
  w'' - c_s^2 \lapthree w - \frac{\theta''}{\theta} w = 0,
\end{equation}
where
\begin{equation}
  \theta = \left(\frac{3}{2}\right)^{1/2} \frac{h}{a} \left[ h^2 - h' + \mathcal{K}\right]^{-1/2}.
\end{equation}
This is a very convenient evolution equation for the metric
perturbations; it is a complete description of the dynamics
of scalar perturbations, under the assumptions made on the stress tensor.

A second form of the evolution equation can be obtained by
introducing a scalar \(u\), which is defined by the relation
\begin{equation}
  \lapthree w = - c_s(\eta)^{-2} \left(\frac{d}{d\eta} + \frac{\theta'}{\theta}\right) c_s(\eta) u.
\end{equation}
Because of the relation between this definition and the differential
operator which appears in the \(w\) equation, it is not difficult
to show that \(u\) satisfies the evolution equation
\begin{equation}
  u'' - c_s(\eta)^2 \lapthree u - \frac{z''}{z} u = 0,
\end{equation}
where
\begin{equation}
  z = \frac{a}{c_s h} \left(h^2 - h' + \mathcal{K}\right)^{1/2}.
\end{equation}
Note also that these definitions relate the quantity \(u/z\) to
the metric perturbation in a simple way,
\begin{equation}
  \lapthree \Phi = - (4\pi G)^{1/2} \frac{h^2 - h' + \mathcal{K}}{h c_s^2} \left(\frac{u}{z}\right)'.
\end{equation}
Recalling that \(\Phi\) has the interpretation of the Newtonian gravitational
potential, this Poisson equation indicates how \(u/z\) acts as a source
for the potential.

Finally we specialize to the situation where the background is dominated
by the dynamics of a scalar field; the only ingredient that we need from this
choice is the stiff nature of a relativistic field, \(c_s = 1\). We also
choose flat spatial sections as appropriate for generic inflationary models,
\(\mathcal{K} = 0\). Since \(u\) has a
canonical Hamiltonian evolution, it is a natural choice for canonical
quantization; the operator is expanded in terms of mode functions
as in Eqn.~(\ref{uexpand}), and the mode functions \(u_k\) are
found to satisfy Eqn.~(\ref{mode}).

The tensor perturbations are expanded in terms of tensorial modes
on the spatial sections,
\begin{equation}
  E_{ij}(\eta ,\mathbf{x}) = \frac{1}{a}
    \int \mathrm{d} \mu (\mathbf{k}) \; 
    v_k (\eta)
    \mathcal{Y}_{ij}(\mathbf{x}),
\end{equation}
where the mode functions \(\mathcal{Y}_{ij}(\mathbf{x})\) have
the properties
\begin{equation}
  (\lapthree + k^2) {\cal Y}_{ij} = 0,\quad
  \gamma^{ij} \mathcal{Y}_{ij} = 0,\quad
  D^{i} {\cal Y}_{ij} = 0.
\end{equation}
Again it is found that the mode functions \(v_k\) satisfy
a simple evolution equation, which is given by Eqn.~(\ref{modev}).

\section{Definition of the Errors}
\label{error}

A key advantage of the uniform approximations presented in
Ref.~\cite{Olver} is the uniform control over the remainder terms
for the approximations. This uniform control is obtained
by carefully separating the dominant influences in the
coefficient functions of the ODE.
Because of this uniform control, this approach is
superior to the earlier results of Langer (for a good description see
Ref.~\cite{BO}). Additionally, Olver constructs higher order
approximations not present in the original work. In this appendix
we review in detail the general error formulae given by Olver. The
errors in Eqns.~(\ref{gensol1}) and (\ref{gensol2}) are bounded by
\begin{eqnarray}\label{err1}
&&\frac{|\epsilon_{2n+1,\lgtr}^{(1)}(b,\xi)|}{M(b^{2/3}\xi)},
\frac{|\partial \epsilon_{2n+1,\lgtr}^{(1)}(b,\xi)/
\partial\xi|}{b^{2/3}N(b^{2/3}\xi)}
\\ \label{err2}
&&\le 2 E^{-1}(b^{2/3}\xi)\exp\left\{ 
\frac{2\lambda{\cal V}_{\xi,\beta}(|\xi|^{1/2}B_0)}{b}
\right\}
\frac{{\cal V}_{\xi,\beta}(|\xi|^{1/2}B_n)}{b^{2n+1}},\nonumber\\
&&\frac{|\epsilon_{2n+1,\lgtr}^{(2)}(b,\xi)|}{M(b^{2/3}\xi)},
\frac{|\partial\epsilon_{2n+1,\lgtr}^{(2)}(b,\xi)/\partial\xi|} 
{u^{2/3}N(b^{2/3}\xi)}
\\
&&\le 2 E(b^{2/3}\xi)\exp\left\{ 
\frac{2\lambda{\cal V}_{\alpha,\xi}(|\xi|^{1/2}B_0)}{b}
\right\}
\frac{{\cal V}_{\alpha,\xi}(|\xi|^{1/2}B_n)}{b^{2n+1}},\nonumber
\end{eqnarray}
where $M(x)$ and $N(x)$ are modulus functions, and $E(x)$ is a weight 
function defined as
\begin{eqnarray}
M(x)&=&\sqrt{2 {\rm Ai}(x){\rm Bi}(x)} \mbox{ for } x\le c,\nonumber\\
M(x)&=&\sqrt{{\rm Ai}^2(x)+{\rm Bi}^2(x)} \mbox{ for } x\ge c,\label{Mx}\\
N(x)&=&\left\{\frac{{\rm Ai'}^2(x){\rm Bi}^2(x)
+{\rm Bi'}^2(x){\rm Ai}^2(x)}
{{\rm Ai}(x){\rm Bi}(x)}\right\}^{1/2} 
\mbox{ for } x\ge c,\nonumber\\
N(x)&=&\left\{{\rm Ai'}^2(x)+{\rm Bi'}^2(x)\right\}^{1/2}\mbox{ for } x\le c,\\
E(x)&=&\sqrt{\frac{{\rm Bi}(x)}{{\rm Ai}(x)}}\mbox{ for } 
c\le x\le \infty,\nonumber\\
E(x)&=&1 \mbox{ for } -\infty\le x\le c,\label{Ex}
\end{eqnarray}
and \(c\simeq -0.36605\).
Some explicit numerical values of these functions are given in
Ref.~\cite{Olver}. The auxiliary quantity $\lambda$ is defined by
\begin{equation}
\lambda=\sup_{(-\infty,\infty)}\left\{\pi|x|^{1/2}M^2(x)\right\}.    
\end{equation}    
A numerical estimate for $\lambda$ is \(\lambda\simeq 1.04\)
\cite{Olver}.  Finally, in Eqns. (\ref{err1}) and (\ref{err2}), we
introduced the total variation of a function over the interval
$(\alpha,\beta)$, ${\cal V}_{\alpha,\beta}(f)$.  The total variation
of a function $f(x)$ over an interval $(\alpha,\beta)$ is the supremum
\begin{equation}
\mathcal{V}_{\alpha,\beta}(f) = \sup_{\{\alpha \le x_0 
< \dots < x_n < \dots \le \beta\}}\sum_{s=0}^{n-1}|f(x_{s+1})-f(x_s)|,
\end{equation}
for unbounded $n$ and all possible subdivisions of the interval,
\(\alpha \le x_0 < \dots < x_n \le \beta\).
In case of a compact  interval $[\alpha,\beta]$ one possible subdivision is
given by $n=1$, $x_0=a$, and $x_1=b$. Hence
\begin{equation}
{\cal V}_{\alpha,\beta}(f)\ge |f(\beta)-f(\alpha)|.
\end{equation}
Equality holds when $f(x)$ is monotonic over $[\alpha,\beta]$.
When $f(x)$ is continuously differentiable in $[\alpha,\beta]$ we have
\begin{equation}
{\cal V}_{\alpha,\beta}(f)=\int_\alpha^\beta|f'(x)|dx.
\end{equation}

\end{appendix}


\begin{thebibliography}{99}

\bibitem{cmbobs}C.B.~Netterfield et al., ApJ {\bf 571}, 604 (2002);
N.W.~Halverson et al., ApJ {\bf 568}, 38 (2002); A.~Benoit et al., AA
{\bf 399}, L19 (2003); T.J.~Pearson et al., ApJ {\bf 591}, 556 (2003);
C.L.~Bennett et al. ApJS {\bf 148}, (2003); Chao-lin Kuo et al., ApJ
{\bf 600}, 32 (2004).

\bibitem{rev} For a review of CMB anisotropies, see W.~Hu and
S.~Dodelson, Ann. Rev. Astron. and Astrophys. {\bf 40}, 171 (2002).

\bibitem{lssobs}e.g., W.J.~Percival et al., MNRAS {\bf 327}, 1297
(2001); K.~Abazajian et al., AJ (to appear), astro-ph/0403325.

\bibitem{infpertorig} V.~Lukash, Pis'ma Zh. Eksp. Teor. Fiz. {\bf 31},
631 (1980) [JETP Lett. {\bf 31}, 596 (1980)]; V.F.~Mukhanov and
G.V.~Chibisov, Pis'ma Zh. Eksp. Teor. Fiz. {\bf 33}, 549 (1981) [JETP
Lett. {\bf 33}, 532 (1981)]; Zh. Eksp. Teor. Phys. {\bf 83}, 475
(1982) [Sov. Phys. JETP {\bf 56}, 258 (1982)]; A.H.~Guth and S.-Y.~Pi,
Phys. Rev. Lett. {\bf 49}, 1110 (1982); S.W.~Hawking, Phys. Lett. B
{\bf 115}, 295 (1982); A.A.~Starobinsky, Phys. Lett. B {\bf 117}, 175
(1982).

\bibitem{givb} J.M.~Bardeen, Phys. Rev. D {\bf 22}, 1882 (1980).

\bibitem{givs} For a readable account of gauge invariant
perturbations, see J.M.~Stewart, Class. Quant. Grav. {\bf 7}, 1169
(1990).

\bibitem{ivrevs1} H.~Kodama and M.~Sasaki, Prog. Th. Phys. Supp. 
{\bf 78}, 1 (1984).

\bibitem{ivrevs2} V.F.~Mukhanov, H.A.~Feldman, and R.H.~Brandenberger,
Phys. Rep. {\bf 215}, 203 (1992).

\bibitem{givinf} J.M.~Bardeen, P.J.~Steinhardt, and M.S.~Turner,
Phys. Rev. D {\bf 28}, 679 (1983).

\bibitem{givvfm} V.F.~Mukhanov, JETP Letters {\bf 41}, 493 (1985); 
Sov. Phys. JETP {\bf 68}, 1297 (1988).

\bibitem{givdhl}D.H.~Lyth, Phys. Rev. D {\bf 31}, 1792 (1985). 

\bibitem{LLKCBA} E.~Lidsey, A.R.~Liddle, E.W.~Kolb, E.J.~Copeland, 
T.~Barreiro, and M.~Abney, Rev. Mod. Phys. {\bf 69}, 373 (1997).

\bibitem{slowrev} A.R.~Liddle and D.H.~Lyth {\em Cosmological
Inflation and Large-Scale Structure} (Cambridge, 2000) and references 
therein.

\bibitem{anisotropy} P.J.E.~Peebles and J.T.~Yu, Ap. J. {\bf 162}, 815
(1970); R.A.~Sunyaev and Ya.B.~Zeldovich, Ap\&SS {\bf 7}, 3 (1970)
treated the case of scalar perturbations, while the tensor case was
considered in Ref.~\cite{gravbg} below.

\bibitem{gravbg} A.A.~Starobinskii, Pis'ma Astron. Zh. {\bf 11}, 323 
(1985) [Sov. Astron. Lett. {\bf 11}, 133 (1985)].

\bibitem{polar} W.~Hu and M.~White, New Astron. {\bf 2}, 323 (1997).

\bibitem{recon} H.M.~Hodges and G.R.~Blumenthal, Phys. Rev. D {\bf
42}, 3329 (1990); E.J.~Copeland, E.W.~Kolb, A.R.~Liddle, and
J.E.~Lidsey, Phys. Rev. D {\bf 48}, 2529 (1993); {\bf 49}, 1840
(1994); S.~Dodelson, W.H.~Kinney, and E.W.~Kolb, Phys. Rev. D {\bf
56}, 3207 (1997); E.J.~Copeland, I.J.~Grivell, E.W.~Kolb, and
A.R.~Liddle, Phys. Rev. D {\bf 58}, 043002 (1998). 

\bibitem{HT} M.B.~Hoffman and M.S.~Turner, Phys. Rev. D {\bf 64}, 
023506 (2001); W.H.~Kinney, Phys. Rev. D {\bf 66}, 083508 (2002); 
W.H.~Kinney and R. Easther, Phys. Rev. D {\bf 67}, 043511 (2003);
S.M.~Leach and A.R.~Liddle, Phys. Rev. D {\bf 68}, 123580 (2003).

\bibitem{paraminf} S.M.~Leach, A.R.~Liddle, J.~Martin, and
D.J.~Schwarz, Phys. Rev. D {\bf 66}, 023515, (2002).  

\bibitem{other} J.~Khoury, B.A.~Ovrut, P.J.~Steinhardt, and N.~Turok,
Phys. Rev. D {\bf 64}, 123522 (2001); P.J.~Steinhardt  and N.~Turok,
Phys. Rev. D {\bf 65}, 126003 (2002).  

\bibitem{hz} E.~Harrison, Phys. Rev. D {\bf 1}, 2726 (1970);
Ya.B.~Zeldovich, MNRAS {\bf 160}, 1 (1972).

\bibitem{cons} R.~Davis {\em et al}, Phys. Rev. Lett. {\bf 69}, 1856 
(1992); D.~Lyth and A.~Liddle, Phys. Lett. B {\bf 291}, 391 (1992).

\bibitem{wms} L.~Wang, V.F.~Mukhanov, and P.J.~Steinhardt,
Phys. Lett. B {\bf 414}, 18 (1997).

\bibitem{ms} J.~Martin and D.J.~Schwarz, Phys. Rev. D {\bf 62}, 103520
(2000).   

\bibitem{srimprov} E.D.~Stewart, Phys. Rev. D {\bf 65}, 103508 (2002).  

\bibitem{schwarzetal} D.J.~Schwarz, C.A.~Terrero-Escalante, and
A.A.~Garcia, Phys. Lett. B {\bf 517}, 243 (2001).

\bibitem{hhjm} S.~Habib, K.~Heitmann, G.~Jungman, and 
C.~Molina-Par\'{\i}s, Phys. Rev. Lett. {\bf 89}, 281301 (2002).

\bibitem{ms2} J.~Martin and D.J.~Schwarz, Phys. Rev. D {\bf 57}, 3302
(1998). 

\bibitem{Olver} F.W.J.~Olver, {\em Asymptotics and Special Functions},
(AKP Classics, Wellesley, MA 1997).

\bibitem{ms03} J.~Martin and D.J.~Schwarz, Phys. Rev. D  {\bf 67},
083512, (2003).

\bibitem{fulling} S.A.~Fulling, Ch. VII, {\em Aspects of Quantum Field
Theory in Curved Space-Time} (Cambridge University Press, New York,
1989).  

\bibitem{shetal} S.~Habib et al., in preparation.

\bibitem{guthpi} A.H.~Guth and S.Y. Pi, Phys. Rev. Lett. {\bf 49},
1110, (1982).

\bibitem{sg} E.D.~Stewart and J.-O. Gong, Phys. Lett. B {\bf 510},
1 (2001).

\bibitem{sl} E.D.~Stewart and D.H.~Lyth, Phys. Lett. B {\bf 302}, 171
(1993). 

\bibitem{LPB} A.R.~Liddle, P.~Parsons, and J.D.~Barrow, Phys. Rev. D
{\bf 50}, 7222 (1994).

\bibitem{CGS}J.~Choe, J.O.~Gong, and E.W.~Stewart, hep-ph/0405155. 

\bibitem{cmbfast} http://www.cmbfast.org; U.~Seljak and
M.~Zaldarriaga, Ap.~J. {\bf 469}, 437 (1996).

\bibitem{BO} C.M.~Bender and S.A.~Orszag {\em Advanced Mathematical
    Methods for Scientists and Engineers} (McGraw-Hill, 1978).

\end{thebibliography}
\end{document}